\documentstyle[12pt,epsfig]{book}
\large
\newcommand{\beq}{\begin{equation}}
\newcommand{\eeq}{\end{equation}}
\newcommand{\bra}{\begin{array}}
\newcommand{\era}{\end{array}}
\newcommand{\be}{\beta}

\let\include\input
\def\be{\begin{equation}}
\def\ee{\end{equation}}
\def\bea{\begin{eqnarray}}
\def\eea{\end{eqnarray}}
\def\nn{\nonumber}
\def\t{\times}
\def\[{\bigl[}
\def\]{\bigr]}
\def\({\bigl(}
\def\){\bigr)}
\def\p{\partial}
\def\o{\over}

\def\a{\alpha}
\begin{document}
\makeatletter \@addtoreset{equation}{section}
\renewcommand{\theequation}{\thesection.\arabic{equation}}
\title{\rightline{\mbox{\small
{GNPHE/0810}}} \textbf{ Introduction \`a  La Th{\'e}orie  } \\
\textbf{Des  Cordes: I }}
\author{ \textbf{Adil Belhaj}\thanks{belhaj@unizar.es} \\
{\small   Centre National de l'Energie, des Sciences et des Techniques Nucl{\'e}aires, Rabat, Morocco,}\\
{\small Groupement National de Physique des Hautes Energies, Si%
\`{e}ge focal: FS, Rabat,  Morocco}} \maketitle
 \vspace{-10cm}


 Ce travail  est une note   de cours sur la th\'eorie
   des supercordes que nous avons  pr\'epar\'es  pour les
    \'etudiants  de doctorats de UFR-PHE Laboratoire de Physique des Hautes
     Enerigies de la facult\'e des Sciences de Rabat (2002, 2003, 2004, 2007, 2008).
     Ce sont des notes pr\'eliminaires  d\'estin\'ees pour des jeunes chercheurs.
\begin{center}
{\bf Remerciement}
\end{center}

Je tiens \`a  exprimer mes profonds remerciements aux chercheurs
suivant pour la discution et l'aide scientifique qu'il m'ont
accord\'e:  M. Asorey, L. Boya,   J. L. Cortes, P. Diaz,  M. P.
Garcia de Moral, C. Gomez, L. Ibanez,  S. Montanez, S.
Randjbar-Daemi,  J. Rasmussen, E. Saidi, A. Sebbar, A. Segui, C.
Vafa, A. Urange.

Enfin, j'adresse mes remerciements \`a ma famille, et tous mes amis au Canada,  Espagne et USA.

\tableofcontents
\chapter*{Introduction G\'en\'erale }
\addcontentsline{toc}{chapter}{Introduction G\'en\'erale }
\pagestyle{myheadings}
\markboth{\underline{\centerline{\textit{\small{Introduction
G\'en\'erale}}}}}{\underline{\centerline{\textit{\small{Introduction
G\'en\'erale}}}}}
 \qquad Un des objectifs de la physique des hautes \'energies est
d'\'elaborer une th\'eorie quantique  saine  capable de d\'ecrire
les quatres interactions fondamentales de l'univers.
Alors que  l'interaction \'electromagn\'etique (s'exercant entre
les  particules charg\'ees), l'interaction faible (responsable des
d\'esint\'egrations   nucl\'eaires)  et l'interaction forte
(permettant  la coh\'esion  des noyaux atomiques)  sont
ad\'equatement  d\'ecrites  dans le cadre   des  mod\`eles  de
grande unification; la gravitation, responsable de l'attraction
mutuelle entre
 les corps mat\'eriels, est cependant trait\'ee \`a
part.  Les mod\`eles de supercordes que nous allons
pr\'esenter  dans ces  notes permettent  de r\'esoudre ce probl\`eme
 au d\'etriment  de la cr\'eation d'autres \cite{GSW,Pol1,Vafa1,Rand1,Na1,GH,Pol2,Kir}.
\par   L'id\'ee de base de la th\'eorie  de supercordes
  consid\'ere qu'au niveau  quantique ($< 10^{-34}$ cm),  les objets
fondamentaux
de la physique  ne sont  plus vus    comme des particules
ponctuelles  de dimension nulle mais plut\^ot  comme  des  objets
\'etendus  de  dimension un  et de
 tension   $T$.
 Il existe  plusieurs
types de cordes:  corde  bosonique (ouverte ou  ferm\'ee) ou
fermionique (ouverte,  ferm\'ee; chirale   et non chirale). A l'\'echelle de Planck,    les cordes
bosoniques ferm\'ees sont  assimil\'ees  \`a  des cercles;  ses
  excitations quantiques contiennent  entre autres  le graviton et un tenseur
antisym\'etrique
   de rang deux jouant un r\^ole crucial en th\'eorie des cordes. Quant aux
       cordes bosoniques ouvertes, elles  sont  assimil\'ees  \`a  des  petits
 segments avec des conditions aux bords
 de Dirichlet   ou   de
Neumann; ses excitations quantiques contiennent naturellement les
champs de jauge
  \cite{GSW,Pol1,Vafa1}. Les deux types  de cordes, ouvertes et ferm\'es, sont
n\'ecessaires
    pour la construction  des mod\`eles quantiques consistents.  \\
\par Lors de son  mouvement classique, la  corde  bosonique (ouverte et
ferm\'ee)
 engendre  dans l'espace-temps une surface bidimensionnelle (surface d'univers).  Par suite, la th\'eorie classique des cordes
peut \^etre vue  comme
 une th\'eorie des champs
bidimensionnelle conforme.  Au niveau quantique, les contraintes
d'invariance conforme
 exigent  que  la dimension  de l'espace-temps soit  \'egale \`a
$D=26$ au lieu  des   (1+3)    habituelles \cite{Poly1,Poly2,Ram,Ven,GSO1,GSO2,BPZ,FQS,FV,NST}.   Ce type de
mod\`ele   est  baptis\'e th\'eorie de corde bosonique.   Cette
th\'eorie n'est cependant  pas  consistante  au niveau quantique
en  raison de:
\begin{itemize}
  \item {Absence des fermions n\'ecessaires  pour d\'ecrire la
mati\`ere.} \item { Pr\'esence  du  tachyon (particule de masse
carr\'ee  n\'egative).}
\end{itemize}
Le premier probl\`eme est r\'esolu par la
consid\'eration des th\'eories de supercordes (cordes
supersym\'etriques)  qui vivent   dans un espace-temps de
dimension r\'eduite \`a $D=10$ et le deuxi\`eme est surmont\'e par
 la projection de    Gliozzi, Sherk et Olive  ({\bf  GSO}) \cite{GSO1,GSO2}.
Les contraintes,  permettant de compenser
 les  anomalies gravitationnelle et de Yang-Mills,
montrent  qu'il  n'existe pas une seule th\'eorie \`a $D=10$, mais
plut\^ot  cinq
  mod\`eles  de supercordes  class\'es  comme suit \cite{Vafa1}
\\  (a) {\it Supercordes ayant une supersym\'etrie d'espace-temps $N=1$ \`a
dix dimensions  comprenant:}
\begin{itemize}
\item{ Supercorde    de type I
avec un groupe de  jauge $SO(32)$} \item{ Supercorde h\'et\'erotique
$SO(32)$}
\item{ Supercorde
h\'et\'erotique   $E_8\times E_8$}
\end {itemize}
(b) { \it Supercordes ayant
une supersym\'etrie d'espace-temps $N=2$:}\begin{itemize}
 \item{ Supercorde
non
chirale  type IIA} \item {Supercorde   chirale  type IIB} \end{itemize}
 Le  spectre des \'etats  non massifs de ces mod\`eles   contient le  dilaton
$\phi$,
  dont  $g_s =e^{\phi}$ est   la constante de  couplage de la th\'eorie, le
graviton
   de spin 2 et   des  tenseurs  antisym\'etriques de  jauge   g\'en\'eralisant
     la  notion  de  potentiel  vecteur  $A_{\mu}$  \`a   des tenseurs
antisym\'etriques
     $A_{\mu_1\ldots\mu_{p+1}}$ \`a $p+1$ indices  (($p+1$)-formes,
$p=1,2,\ldots$).
      Ces tenseurs  se couplent  \`a des objets \'etendus  appel\'es
$p$-branes qui
       g\'en\'eralisent  la particule ponctuelle $(p=0)$ et la  corde  $(p=1)$.
          Les $p$-branes  sont  infiniment lourdes \`a faible couplage et
occupent
            des hypersufaces de $(p+1)$ dimensions  dans l'espace-temps \`a dix
              dimensions. Il existe diff\'erents    types de branes:  {\bf
NS-NS}
                branes   qui sont charg\'ees  sous  l'action des tenseurs du
secteur
                 {\bf NS-NS} et  D-branes charg\'ees sous les tenseurs  {\bf
R-R} \cite{ON,Gren1,Di,Anto,Malda1,Kach,Malda2,Malda3,Kuta}. Ces derni\`eres sont stables   et jouent un r\^ole
 important dans la description du r\'egime non perturbatif  des mod\`eles de
supercordes \cite{Vafa1}.\par
  La  th\'eorie des supercordes  peut \^etre \'egalement d\'efinie dans des
dimensions
   inf\'erieures  \`a $D=10$  en utilisant la m\'ethode de compactification
\cite{Vafa1,CHSW}.
   Cette derni\`ere suppose que certaines   des dix dimensions  sont compactes
 et non observables  \`a notre \'echelle.
Les compactifications les plus \'etudi\'ees en  th\'eorie   des supercordes
              sont:  \begin {itemize}  \item { Compactification toroidale qui
pr\'eserve   toutes  les  supercharges initiales  \cite{NSW}.}  \item{
  Compactification   sur des  vari\'et\'es  de Calabi-Yau pr\'eservant un
certain nombre de supercharges \`a dix dimensions \cite{Yau,Gren2,Thei,Dou}.}
\end {itemize}
 Ces derni\`eres  restent   les candidats  les  plus probables   pour
connecter les mod\`eles de  supercordes  au notre  monde  r\'eel.\par
Puisqu'il existe cinq types de mod\`eles de supercordes
consistents
 \`a dix dimensions, nous allons  se trouver  avec un grand nombre de
th\'eories
  de supercordes \`a des dimensions inf\'erieures.   Ces mod\`eles
       sont diff\'erents de ceux des supercordes fondamentales \`a dix
dimensions
        dont les  \'etats du vide  sont d\'etermin\'es par les valeurs moyennes
           des champs scalaires  (appel\'ees aussi modules )
interpr\'et\'ees comme
            les coordonn\'ees locales d'un  espace  communement  appel\'e
espace des modules. \par
              La compactification  est   une m\'ethode   permettant   non
seulement
                  de  r\'eduire  la dimension de l'espace-temps  mais
                        offre  aussi  des    possibilit\'es  de  connecter les
 diff\'erents
                           mod\`eles de supercordes  dans  les  dimensions
inf\'erieures  par
                            le biais des    sym\'etries de dualit\'es. Des
progr\`es r\'ecents
                             conduisent \`a penser que ces
diff\'erentes th\'eories ne sont en fait  que des cas limites
d'une th\'eorie unique,  appel\'ee th\'eorie-M (M pour  m\`ere)
\cite{Witten1}.  Les sym\'etries  de dualit\'es relient des r\'egimes
de couplage diff\'erents g\'en\'eralisant    ainsi  la dualit\'e
\'electrique-magn\'etique
 observ\'e par le pass\'e dans le cadre
des th\'eories de jauge \`a quatre dimensions. L'exemple le plus
simple de sym\'etrie de dualit\'e en th\'eorie  des supercordes
est  nulle doute la   dualit\'e-S  de la th\'eorie   IIB  \`a dix
dimensions  \'echangeant
  la constante de couplage   $g_{IIB}$ en  $1 \over g_{IIB}$
 \cite{Schwarz}.  Un autre exemple est donn\'e par   les supercordes de type I
  $ SO(32)$ et h\'et\'erotique  $ SO(32)$  qui sont \'echang\'ees  moyennant
la substitution    $  g_{het}={1 \over {g_I}}$.   La supercorde
h\'et\'erotique peut \^etre vue alors comme le soliton de la
supercorde
 type I  \`a dix dimensions.  Ces arguments conduisent \`a consid\'erer   les
mod\`eles de type I $ SO(32)$ et la supercorde h\'et\'erotique $
SO(32)$ comme  deux d\'eveloppements   perturbatifs extr\^emes,
l'un \`a $g_I$ faible et  l'autre \`a   $ 1\over g_I$    faible
d'une m\^eme  th\'eorie d\'efinie en toute valeur de  $ g_I$.
\par Les sym\'etries  de dualit\'es, qui   sont  ad\'equatement   d\'ecrite
  par la sym\'etrie miroir introduite  dans  \cite{Vafa1},  existent   \'egalement
    pour les mod\`eles de supercordres  dans des dimensions d'espace-temps
inf\'erieures
     \`a dix. A neuf dimensions par exemple,  les espaces des modules
 des  supercordes h\'et\'erotiques  $ E_8\times
E_8$ et $SO(32)$ sont  perturbativement \'equivalents par  la
dualit\'e-T  apr\`es compactification sur un cercle. De m\^eme les
mod\`eles  de supercordes  de type IIA et type IIB sont isomorphes
sous la dualit\'e-T \cite{Vafa1}. En pratique,
 il se trouve que  le cas le plus important  de
 dualit\'e corde-corde est celui  reliant la supercorde  IIA sur une
vari\'et\'e
    K3  et   la    supercorde h\'et\'erotique   sur un tore $T^4$ \cite{Vafa1,Witten1,HT,DKV,KLMVN,KKV}.
     Ces  deux mod\`eles
      pr\'esentent  le m\^eme espace des modules
       $  {\bf R^+ }\times {SO(20,4,{\bf R})\over  SO(20)\times SO(4)}$
       d\'ecrivant   le dilaton  $({\bf R^+ })$ et  les  modules
        du  r\'eseau  pair  auto-dual de Narain  $\Gamma^{20,4}$
        d\'efinissant la compactification  de supercorde h\'et\'erotique sur
$T^4$
         ou encore  la compactification
            de la th\'eorie de supercorde    de type IIA sur
             K3.   Les \'etats  non massifs  du
               spectre   perturbatif    correspondant au r\'eseau
                de Narain $\Gamma^{20,4}$  sont identifi\'es avec  les
D-2branes  de IIA sur K3
                 enroul\'ees  sur  les   2-cycles  d'aire nulles de
                 l'homologie de K3.  Autrement dit la sym\'etrie de jauge
                  perturbative des  supercordes
                   h\'et\'erotiques sur $T^4$  est  identifi\'ee  avec les
                    singularit\'es  ADE  de K3 \cite{KLMVN}. Ainsi \`a  six
dimensions,
                     l'\'etude  de la sym\'etrie de jauge non ab\'elienne
                      dans le mod\`ele de supercorde type IIA
                       aux  faibles \'energies est rattach\'ee \`a
                        l'\'etude des singularit\'es  de
                         K3 en pr\'esence des D-2branes   enroul\'ees
                           sur les 2-cycles de K3.  Cette id\'ee  a
                            \'et\'e d\'evelopp\'ee      C. Vafa  et  al   pour \'etudier la limite des
th\'eories des champs \`a six et \`a  quatre  dimensions obtenues
 \`a partir de la  supercorde IIA et de la  th\'eorie-F \cite{Vafaf}. Cette \'etude a
donn\'e naissance  \`a l'approche  dite  {\it ing\'enierie
                               g\'eom\'etrique} (geometric engineering)  des
th\'eories  des champs quantiques \cite{KKV,KMV}.  Cette
construction va  au del\`a du  c\'el\`ebre  mod\`ele
de Seiberg-Witten des th\'eories de jauge
           supersym\'etriques $N=2$ \`a quatre dimensions \cite{SW1,SW2}  et
             offre une alternative \`a l'approche de Hanany-Witten des
th\'eories
              de jauge bas\'ee  sur l'introduction des D-branes
\cite{HW,Witten2}.
               Le succ\`e de  l'ing\'enierie
                               g\'eom\'etrique   est  essentiellement  du
             aux m\'ethodes sophistiqu\'ees de
              la g\'eom\'etrie torique des vari\'et\'es
              de Calabi-Yau   locales et de  la sym\'etrie miroir. L'id\'ee
principale de l'ing{\'e}nierie g{\'e}om{\'e}trique des th\'eories
 quantiques
  des champs  \`a partir   des mod\`eles  de supercordes  est de partir
    de   la supercorde de type
                 IIA, en pr\'esence des D2-branes, compactifi\'ee  sur des
vari{\'e}t{\'e}s de Calabi-Yau locales  \` a trois  dimensions  complexes.
Ceci donne
                 naissance,  \`a faible \'energie,  \`a  des th{\'e}ories des
champs
                    supersym{\'e}triques $N=2$ {\`a} quatre  dimensions dont
   l'espace des modules physique $\cal M$ contient
                     l'espace  des modules  g\'eom\'etriques    de Calabi-Yau
\`a trois
                     dimensions complexes. En
fait,
                      la compactification de la th{\'e}orie  IIA sur  K3,
                       en pr{\'e}sence de D2-branes,  conduit {\`a} des
th{\'e}ories
                        supersym{\'e}triques $ N=2$ {\`a} six  dimensions; une
 deuxi\`eme
                          compactification  sur un espace de dimension 2,
conduira  alors
                            {\`a} des mod\`eles de type IIA {\`a}  quatre
dimensions.
\par

Le   but    central de  ces  notes  est de donner   une revue sur
  les recents d\'eveloppements  de  la  th\'eorie
  supercordes;  tout particuli\`erement   leurs  spectres  d'\'etats non
  massifs, classification des supercordes, compactification et
  dualiti\'es.\\
  Dans le  premi\`er chapitre  nous rappelons la th\'eorie   quantique des
cordes
   bosoniques (ouvertes et ferm\'ees)  vivant \`a 26 dimensions.   Par suite
nous
    introduisons dans le chapitre deux la th\'eorie des supercordes
g\'en\'eralisant la corde bosonique
     vers une th\'eorie plus saine   \`a  $D=10$. En utilisant la projection
de  Gliozzi,
      Sherk et Olive  ({\bf GSO}),  nous construisons le  spectre d'\'etats
non massifs  de
      la th\'eorie des supercordes \`a  $D=10$. Finalement nous donnons une
classification
       des cinq mod\`eles  de supercordes  ainsi  que leurs spectres \`a
$D=10$.\\
  Dans le  chapitre 3,  nous \'etudions les  diff\'erentes
   compactifications des mod\`eles de supercordes \`a  $D=10$  vers des
    dimensions inf\'erieures.  En  premier temps, nous examinons     trois
exemples de  compactification
       des supercordes:
 La compactification toroidale sur un  tore  $T^d$, \ la  compactification sur
 K3 de groupe
   d'holonomie $SU(2)$ et
la  compactification sur des vari{\'e}t{\'e}s de Calabi-Yau \`a
trois dimensions  complexes.
 Afin de mieux illustrer la compactification  sur $T^d$, nous commen\c cons
   tout d'abord par la compactification    des supercordes  sur  un cercle
$S^1$.  Ensuite nous donnons les
      compactifications des cinq   mod\`eles des supercordes sur  la
vari\'et\'e K3  (vue comme orbifold)  ainsi  que  sur des vari\'et\'es complexes   de Calabi-Yau
       \`a trois dimensions.   Apr\`es, nous montrons  comment
        peut-on  construire  les mod\`eles duaux
        en utilisant  la compactification.   Cette \'etude    montre que le
r\'egime
        non perturbatif des mod\`eles de supercordes est   bas\'e sur
l'\'etude des
 {\it D- branes} \cite{Vafa1} dont on rappelle  la physique dans le chapitre
quatre.   Nous  profitons de cette
  \'etude  pour  prouver les sym\'etries de dualit\'es   des mod\`eles de
superordres en
   utilisant comme argument de d\'epart  la dualit\'e  entre la supercorde
h\'et\'erotique
    sur le tore $ T^4$  et    la  supercorde de type IIA sur K3.  Nous
montrons,  \`a partir
     de l'aspect singulier  de K3,  comment extraire  de mani\`ere
g\'eom\'etrique  les
      sym\'etries de jauge non ab\'elienne dans le mod\`ele de supercorde IIA.
Apr\`es  nous etendons cette dualit\'e\`a quatre dimensions. \\

\chapter{   Th\'eorie de la  corde  Bosonique}
\pagestyle{myheadings}
\markboth{\underline{\centerline{\textit{\small{ Th\'eorie de la corde
Bosonique}}}}}{\underline{\centerline{\textit{\small{ Th\'eorie de la  corde
Bosonique}}}}} Couronn{\`e}e  de succ{\'e}s pour la description
des trois interactions
 {\'e}lectromagn{\'e}tique, forte et faible dans le cadre du mod\`ele standard,
  la th{\'e}orie  quantique  des champs  de  particules \'el\'ementaires
devrait
     cependant reconnaitre son {\'e}chec {\`a} d{\'e}crire la force
     gravitationnelle
      au niveau quantique.  La th{\'e}orie des  supercordes apparait {\`a}
l'heure actuelle comme le seul candidat convenable {\`a} l'unification quantique des
quatre  interactions
        fondamentales de la nature.   La th{\'e}orie des  supercordes    a
\'et\'e
          introduite dans les ann{\'e}es 1970, mais son {\'e}tude  r{\'e}elle
n'a commenc{\'e}
            qu'\`a partir des ann{\'e}es 1980.  Classiquement  cette
th\'eorie d\'ecrit
              la dynamique relativiste d'un objet unidimensionnel balayant une
surface d'univers
               au cours de sa propagation dans l'espace-temps;  tout comme la
trajectoire d'une
                 particule \'el\'ementaire    qui  est une ligne d'univers.
De cette
                  propri\'et\'e, il   d\'ecoule  que la  th\'eorie  des
supercordes peut \^etre
                   vue comme  une  th\'eorie   des champs bidimensionnelle
conforme.  Il existe  deux types de cordes:  la corde
ouverte et
                      la corde ferm{\'e}e  vues, \`a petite \'echelle, comme
un interval ou un cercle respectivement.  Les degr\'es de libert\'e (ddl)
des secteurs gauche (L)  et  droit (R) de la corde  ferm\'ee sont
essentiellement ind\'ependants.  La corde ouverte poss\`ede alors la  moiti\'e
 des (ddl) de la corde ferm{\'e}e \`a cause des conditions au bords. \par
L'\'etude quantique     de la th{\'e}orie   des  cordes r\'ev\`ele
la distinction de
 deux types de cordes:  la corde bosonique et la corde fermionique ou
supercorde. Cette
  derni\`ere admet une   surface d'univers supersym\'etrique.   Ceci signifie
qu'en plus
     des coordonn\'ees bosoniques,  nous avons aussi  des coordonn{\'e}es
fermioniques
     r\'ealisant la supersym\'etrie de la surface d'univers.    Les
contraintes d'invariance
      conforme de la  th{\'e}orie  quantique des cordes  exigent  que la
dimension critique de
       l'espace-temps soit  $D = 26$  pour la corde bosonique et  $D = 10$
pour la supercorde.
A premi\`ere vue,  ces dimensions d'espace-temps semblent
\'etranges;
 elles sont cependant le prix \`a payer pour avoir une th\'eorie quantique
  consistante  capable de d\'ecrire l'unification des quatre interactions
fondamentales
    de la nature.   Dans chapitre 3, nous allons voir comment r\'eduire ces
dimensions
     \`a la dimension usuelle de l'espace-temps   \`a notre \'echelle.  Outre
la dimension 26,
       la th\'eorie quantique de la corde bosonique admet d'autres probl\`emes
   dus \`a
        la pr{\'e}sence d'une particule de masse carr{\'e}e n{\'e}gative (le
tachyon)
         ainsi que de l'absence de degr{\'e}s de libert{\'e} fermioniques dans
l'espace-temps.
Ces deux probl\`emes sont absents en th\'eorie quantique des
supercordes \`a dix
  dimensions. C'est la raison  qui nous a pouss\'e \`a   concentrer
     notre attention  sur  l'\'etude des mod\`eles de supercordes. Cependant
      il faudrait noter que malgr\'e ses d\'efauts, l'\'etude de la th\'eorie
quantique
       de la  corde bosonique reste toujours un sujet  de  rechereche
d'actualit\'e;
           en particulier en connection avec  les r\'ecents  d\'eveloppements
sur
            la conjecture de Sen concernant le probl\`eme de condensation des
tachyons \cite{Sentac,Sentac1}. \\
La pr\'esentation de ce chapitre est comme suit: Dans les sections
1 et 2, nous introduisons
  les grandes lignes de la th\'eorie quantique bosonique.
   L\`a aussi nous  donnons seulement  les r\'esultats  fondamentaux.
Plus  de d\'etails pourront  \^etre trouv\'es  dans   \cite{GSW,Pol1,Vafa1}.
\section { Corde  bosonique classique }
\subsection{Approche  de Polyakov}
 L'id\'ee  centrale  de la  th\'eorie  des cordes est  de  consid\'erer
  les objets fondamentaux de la physique classique  non plus  comme des
particules
   ponctuelles (de dimension 0) mais plut\^ot comme des objets de dimension
1, dot\'es
    d'une longeur tr\`es petite.  Les diff\'erentes particules
\'el\'ementaires que nous
     connaissons  apparaitraient alors comme diff\'erents modes de vibration
de la  corde.
       Le but essentiel de la  th{\'e}orie des cordes est alors  la
construction d'une
        m{\'e}canique quantique  relativiste des objets  {\'e}tendus ayant une
structure
         interne de  dimension 1.  Pour mieux  illustrer   l'id\'ee,  il est
int\'eressant
           de rappeler  tout d'abord  l'action  classique d'une particule
libre ponctuelle
            relativiste $X^{\mu}(\tau)$,  de masse $m$  \cite{Vafa1,Kir}:
\begin{equation}
S=-m\int d\tau\sqrt{{dX^\mu\over{d\tau}}{dX^\nu\over{d\tau}}},
\end{equation}
o\`u $\tau$ est le temps propre. Par analogie, l'action  classique
de  la corde d\'ecrite   par  $X^{\mu}(\tau,\sigma)$,
 o\`u  $\sigma$ est l'extension unidimensionnelle, est donn\'ee par
\begin{equation}
S=-T\int d\tau d\sigma\sqrt{|det\;
\partial_{\alpha}X^{\mu}\partial^{\alpha}X_{\mu}|},
\end{equation}
o{\`u} $T={1\o 2\pi\a'}$ est   la tension de la corde et $\alpha'$
est un param\`etre de  dimension $[m]^{-2}$.  $(\tau,\sigma)$ sont
les coordonn\'ees  de la surface d'univers  et  $(\alpha,\beta)$
d{\'e}signent  les directions $(\tau$ et $\sigma)$.  Equation
(1.1.2) dite  l'action  Nambu - Goto,   est une  action   tr{\`e}s
difficile \`a manier  \`a cause de la  racine  carr{\'e}e.  Pour d\'epasser cette difficult\'e technique, Poyakov
a introduit une autre action,  classiquement  {\'e}quivalente
{\`a} l'action (1.1.2),   poss{\`e}dant  les m{\^e}mes
propri{\'e}t{\'e}s  de sym{\'e}trie mais a l'avantage d'{\^e}tre
quadratique  en  les coordonn\'ees $X^\mu(\tau,\sigma)$
\cite{Poly1,Poly2}:
\begin{equation}
S=-{T\over2}\int  d\tau
d\sigma{\sqrt{-h}}h^{\alpha\beta}\partial_{\alpha}X^{\mu}\partial_{\beta}X_{\mu}.
\end{equation}
 En plus  de  $X^\mu(\tau,\sigma)$, cette action  contient un  champ
auxiliaire
  $h_{\alpha\beta}(\tau,\sigma)$  d\'ecrivant la    m{\'e}trique   de la
surface d'univers.
 $h$ est le d{\'e}terminant de la matrice $h_{\alpha\beta}(\tau,\sigma)$ et
$h^{\alpha\beta}(\tau,\sigma)$ est l'inverse de
$h_{\alpha\beta}(\tau,\sigma)$.
L' action (1.1.3) admet plusieurs  sym\'etries locales:
\begin{itemize}
  \item { Le groupe de  Poincar{\'e} dont les \'el\'ements $(\Lambda^\mu_\nu
,a^\nu )$ agissant comme:
\begin{equation}
X^{\mu} \to \Lambda^\mu_\nu X^\nu+a^\nu
\end{equation}}
  \item { Le groupe de  la reparam\'etrisation (diff\'eomorphisme) de la
surface d'univers
\begin{eqnarray}
\sigma \to \sigma'(\sigma,\tau)\\ \tau \to \tau'(\sigma,\tau)
\end{eqnarray} }
\item {La sym{\'e}trie de Weyl:
\begin{equation}
h_{\alpha\beta}\rightarrow  e^{f(\tau ,\sigma)}h_{\alpha\beta}.
\end{equation}}
\end{itemize}
 Les variations  de l'action de Polyakov par rapport \`a la m\'etrique
$h^{\alpha\beta}$
  et les  champs $X^{\mu}$ conduisent aux \'equations suivantes
\be
 T_{\a \beta}= {\sqrt {-h}
}(\partial_{\alpha}X^{\mu}\partial_{\beta}X_{\mu})-{1 \over 2} h^{\alpha\beta}h^{\lambda\sigma}\partial_{\lambda}X^{\mu}\partial_{\sigma}X_{\mu}
\ee
 et
\be
\p_\a({\sqrt h}h^{\alpha\beta}\partial_{\beta}X^{\mu})=0, \ee o\`u
$T^{\alpha\beta}$ est le tenseur \'energie-implusion.  C'est un
tenseur
  sym\'etrique conserv\'e et de trace nulle, ce qui montre que  (1.1.3) est
une th\'eorie
    des champs conforme \`a deux dimensions.
En  utilisant l'invariance sous les diff\'eomorphismes de la
surface d'univers,
 nous pouvons toujours ramener  la m\'etrique $h^{\alpha\beta}$   {\`a} la
forme diagonale
  et pseudo-euclidienne sur la surface d'univers
\begin{equation}
h_{\alpha\beta}=\eta_{\alpha\beta}=\left(\matrix{ -1&0\cr
0&1\cr}\right).
\end{equation}
Ce choix de m{\'e}trique, d{\'e}finissant   la jauge conforme,
ramene   l'action (1.1.3)  \`a
\begin{equation}
S={T\over2}\int d\tau d\sigma \eta^{\alpha\beta}
\partial_{\alpha}X^\mu\partial_\beta X_{\mu}.
\end{equation}
 Alors que  les \'equations (1.1.8-9)  deviennent respectivement  comme suit
\be
 T_{\a \beta}=\partial_{\a}X^{\mu}\partial_{\beta}X_{\mu},\nn
\ee
\begin{equation}
\partial_{\alpha}\partial^{\alpha}X^\mu(\tau,
\sigma)=({\partial^2\over{\partial \sigma^2}}-{\partial^2\over{\partial
\tau^2}})X^\mu(\tau, \sigma)=0.
\end{equation}
 La sym{\'e}trie classique  de Weyl  est bris{\'e}e  au niveau
quantique conduisant alors
  \`a  une anomalie.  Cette anomalie s'annule uniquement si la  dimension de
l'espace-temps est  $26$, (c'est  la
 dimension critique de la th{\'e}orie des cordes bosoniques)\cite{GSW,Pol1,Vafa1}.\\
 Pour fixer les id\'ees nous prenons la dimension de l'espace-temps $(
X^\mu,\mu=0,1,\ldots,D-1)$ \'egale  \`a 26.
La solution g{\'e}n{\'e}rale  de l'equation  de mouvement (1.1.13) s'{\'e}crit
g{\'e}n{\'e}ralement   comme  la somme de deux composantes
arbitraires  $X^\mu_L$ et $X^\mu_R$ d\'ecrivant respectivement les
vibrations gauches  $ L$ ( $L$ pour left) et droites $R$ ($R$ pour
right)
\begin{equation}
X^\mu(\tau, \sigma)= X^\mu_L(\tau- \sigma)+X^\mu_R(\tau+\sigma).
\end{equation}
Ces modes sont ind\'ependants  pour le cas de la corde ferm\'ee
mais pas pour  la corde ouverte comme nous le verrons dans le
paragraphe suivant.
\subsection{Corde ouverte}
 Dans le cas de la corde ouverte, o\`u la coordonn\'ee   $\sigma$  varie
entre 0 et  $\pi$,
 les champs  $X^\mu(\tau,\sigma)$  satisfont  des conditions aux bords  et
par cons\'equent
 les d{\'e}gr{\'e}s de libert\'e (ddl)  du  secteur gauche et droite sont
li\'es. Ce qui restreint
  le nombre de d{\'e}gr{\'e}s de libert\'e  (ddl )  de la th\'eorie de corde
ouverte  \`a  moiti\`e.
On  distingue ainsi deux types d'extr{\'e}mit{\'e}s pour les cordes ouvertes:\\
1- Celles qui  ob\'eissent {\`a} la condition aux limites de {\bf
Neumann}
\begin{equation}
\partial_\sigma X^\mu(\tau, \sigma=0,\pi)=0.
\end{equation}
2- Celles qui restent fixes dans un hyperplan et elles r\'epondent
{\`a} la condition aux limites de {\bf  Dirichlet}
\begin{equation}
\label{equation}
\partial_\tau X^\mu(\tau, \sigma=0,\pi) =0.
\end{equation}
La  solution g{\'e}n{\'e}rale  de l'\'equation de mouvement pour
la corde  bosonique ouverte  est donn\'ee  par:
\begin{equation}
X^\mu(\tau, \sigma)=x^\mu+\ell^2 p^\mu \tau+i\ell \sum_{n\neq 0}
{1\over n}\alpha^\mu_n e^{-in\tau}\cos n\sigma,
\end{equation}
o\`u $\ell$ est  une constante arbitraire avec une dimension  de
longeur.  $x^\mu$ et  $p^\mu$  sont deux constantes
d'int{\'e}gration qui correspondent   \`a  la coordonn{\'e}e  de
centre de masse et  {\`a} l'impulsion totale de la corde
respectivement. Les  $\alpha^\mu_n$ sont  les  modes
d'oscillations de la corde satisfaisant  la  condition de
r\'ealit\'e:
\begin{equation}
\alpha^\mu_{-n}= (\alpha^\mu_{n})^*.
\end{equation}
\subsection{Corde ferm{\'e}e}
La  corde  ferm{\'e}e  n'a pas d'extr{\'e}mit{\'e} et  peut
\^etre vue comme  un  cercle.   Ainsi   les coordonn\'ees  $X^\mu
(\tau, \sigma)$  doivent v\'erifier   les conditions  de
p\'eriodicit\'e suivantes:
\begin{equation}
X^\mu(\tau,\sigma)=X^\mu(\tau,\sigma+2\pi).
\end{equation}
 Dans ce cas,  les degr\'es de lib\'eret\'e    des secteurs droit  $X^\mu_R$
et  gauche $X^\mu_L$ sont  ind{\'e}pendants   et ont pour forme explicite
\begin{equation}
\begin{array}{lcr}
X^\mu_L(\tau-\sigma)= {x^\mu \over 2}+{1\over2}\ell^2
p^\mu(\tau-\sigma)+{i\over2} \ell\sum  \limits _{n\neq 0}  {1\over
n} \alpha^\mu _n e^{-in(\tau-\sigma)}\\
X^\mu_R(\tau+\sigma)= { x^\mu \over 2}+{1\over2}\ell^2
p^\mu(\tau+\sigma)+{i\over2}\ell \sum \limits_{n\neq 0}  {1\over
n} \tilde \alpha^\mu _n e^{-in(\tau+\sigma)}.
\end{array}
\end{equation}
Notons que   nous avons ici deux ensembles des coordonn{\'e}es
d'oscillations $\alpha_n^\mu$ et $ \tilde \alpha_n^\mu$
satisfaisant:
\begin{equation}
\alpha^\mu_{-n}= (\alpha^\mu_{n})^*,\;\;\;\tilde\alpha^\mu_{-n}=
(\tilde \alpha^\mu_{n})^*.
\end{equation}
Les r{\`e}gles de commutation canonique des champs
$X^\mu(\tau,\sigma)$ sont:
\begin{equation}
\begin{array}{lcr}
\{P^\mu(\tau, \sigma), X^\nu(\tau, \sigma')\}={1\o
T}\eta^{\mu\nu}\delta(\sigma-\sigma')\\ \{P^\mu(\tau,
\sigma),P^\nu(\tau, \sigma)\} =\{X^\nu(\tau, \sigma),X^\nu(\tau,
\sigma)\}=0,
\end{array}
\end{equation}
 o\`u $ P^\mu =T {\partial \over \partial \tau} X^\mu (\tau, \sigma)$.
Partant des {\'e}quations (1.1.20,22),
 nous obtenons les \'equations suivantes
\be
\{\a^{\mu}_{m},\a^{\nu}_{n}\}=\{{\tilde \a}^{\mu}_{m},{\tilde
\a}^{\nu}_{n}\}=-im\delta_{m+n,0}\eta^{\mu\nu} \ee
\be
\{{\tilde \a}^{\mu}_{m},{\a}^{\nu}_{n}\}=0 \ee
\be
\{{x}^{\mu},{p}^{\nu}\}=\eta^{\mu\nu}. \ee
 De m\^eme l'Hamiltonien  pour  la corde ferm\'ee
\be
{\cal H}= {T\o2 }\int d\sigma ({\p X \o {\p \tau}}^2+{\p X \o {\p
\sigma}}^2), \ee
 peut {\^e}tre r\'ecrit  comme suit
\be
{\cal H}={1\o2} \sum\limits_{n\in {\bf Z}}(\a_{-n}\a_n+{\tilde
\a}_{-n}{\tilde \a}_n). \ee Tandis que pour la corde ouverte, nous
avons la forme suivante
\be
{\cal H}={1\o 2} \sum \limits_{n \in {\bf Z}}\a_{-n}\a_n. \ee
 L'\'etape suivante  consiste \`a quantifier la th\'eorie de la corde
bosonique critique.
 Puisqu'il existe plusieurs m\'ethodes de quantification covariantes et non
covariantes,
 nous allons nous  concentrer  sur  l'\'etude de la quantification covariante
canonique d\'eduite
imm\'ediatement \`a partir des  \'equations pr\'ec\'edentes en
rempla\c cant   les crochets de poisson $\{,\}$ par  les
commutateurs
\be
\{,\}\to-i\[,\], \ee
\section{Quantification canonique de la corde bosonique}
\subsection{Corde ouverte quantique }
Nous \'etudions   ci-dessous la quantification  de la  corde
ouverte,  ou \'egalement  d'un  secteur  d'oscillation  disons le
secteur gauche de la corde  ferm\'ee.  Les
r\'esultats obtenus s'\'etendent   imm\'ediatement au secteur
droit.
 A partir des {\'e}quations (1.1.23-25),  nous trouvons  les r\'egles de
commutations suivantes:
\begin{equation}
[x^\mu,p^\nu]=i\eta^{\mu\nu}
\end{equation}
\begin{equation}
[\alpha^\mu_m,\alpha^\nu_n]=m\delta_{m+n,0}\eta^{\mu\nu}, \quad
\alpha^+_n=\alpha_{-n}.
\end{equation}
Il  d\'ecoule de  ces \'equations  que $\alpha_{-n}$ et
$\alpha_{n}$ $  (n>0)$  peuvent \^etre vus comme   les
op{\'e}rateurs de cr{\'e}ation et d'annihilation qui  agissent sur
le vide de l'espace de  Fock $ |0,p^\mu>$  d'impulsion $p^\mu$
comme
\begin{equation}
\begin{array}{lcr}
\alpha ^\mu_n|0, p^\mu >=0, \quad n>0\\ p^\mu |0,p^\mu >= p^\mu
|0,p^\mu >.
\end{array}
\end{equation}
Le mode  z{\'e}ro $\alpha^\mu_{0}= {\alpha^ \mu
_{0}}^+={p^\mu\over2}$ correspond {\`a} l'impulsion totale  de la
corde ouverte. L'invariance de jauge  sous les diff\'eomorphismes
de  la  surface d'univers imposant  l'annulation du tenseur
{\'e}nergie-impulsion sur la surface d'univers se manifeste au
niveau quantique  par les conditions suivantes
\begin{equation}
\begin{array}{lcr}
L_n|0,p^\mu >=0, \quad n>0,\\ L_0 |0,p^\mu >=a |0,p^\mu >,
\end{array}
\end{equation}
o{\`u}  $L_n$ sont les modes de Fourier du tenseur \'energie-impulsion gauche
donn\'es par
\begin{equation}
L_m={1\over 2}\sum\limits_
{n=-\infty}^{+\infty}\alpha_{m-n}\alpha_n;\quad n \neq 0.
\end{equation}

et v\'erifient l'ag\`ebre de Virasoro

\begin{equation}
[L_m,L_n]= (m-n)L_{m+n}+\frac{c}{12}m(m^2-1) \delta_{m+n},
\end{equation}
ou $c$  est la charge centrale.
$L_0$ est donn\'e par
\begin{equation}
L_0={\cal H}=-{\alpha' \over4} p^2+\sum \limits_ {n=1 }^{+\infty}
\alpha_{-n}\alpha_n-a,
\end{equation}
o{\`u}  $a$   est une  constante  dite d'ordre normale qui est fix{\'e}e
{\`a} la valeur $a=1$
en exigeant l'unitarit\'e de la th\'eorie quantique, c.\`a.d  l'absence des
\'etats de norme n\'egative.
 En particulier, l'annulation de $L_0$ implique  que la formule de masse
s'\'ecrit sous la forme
\bea
 M^2&=&{2\over \alpha'}(N_L-1)=-p^\mu p_\mu  \nn\\
&=&{2\over \alpha'}(\sum \limits_
{n=1}^{+\infty}\alpha_{-n}\alpha_{n}-1), \eea
 o{\`u}  $N_L=\sum \limits_ {n=1}^{+\infty}\alpha_{-n}\alpha_n$ d{\'e}signe
 les nombres d'oscillations  gauche et $ \alpha' =\ell^2$  est reli\'e \` a la
  tension de la corde ouverte. \subsection{Spectre de masse des \'etats de  la
corde ouverte}
 Soit   un \'etat physique $|\psi>$. Consid\'erons  les
    \'equations  (1.2.7).
D\'eterminons le spectre de masse des \'etats de la corde
bosonique:\\ - L'\'etat fondamental $|0, p>$  a une masse
n\'egative $-2$. \\-Le premier \'etat excit\'e $\alpha^\mu
_{-1}|0, p>$ , construit par l'action de l'op\'erateur
decr\'eation $\alpha^\mu_{-1}$ sur l'\'etat fondamental $|0, p>$,
poss\`ede une masse nulle. Afin d'\'eliminer les \'etats de norme
n\'egative, nous introduisons les $D$ vecteurs de polarisations
$\zeta_\mu^{(\lambda)}$ $\mu, \lambda= 0,\ldots, D-1$, tels que
${(\zeta_\mu^{(0)})}^2=-1$ et ${(\zeta_\mu^{(i)})}^2=1$ , $i =
1,\ldots, D-1$.  Les premiers \'etats excit\'es
 sont de normes \'egales \`a  ${(\zeta_\mu^{(\lambda)})}^2$ .
\subsection{Corde ferm\'ee quantique }
  Concernant la th{\'e}orie des cordes ferm{\'e}es,   les r{\'e}sultats sont
l\'eg\`erement diff{\'e}rents \`a
cause de la  pr\'esence simultan\'ee des deux ensembles des oscillations.
Rappelons  que   les excitations quantiques
 de la corde ferm{\' e}e sont d\'ecrites par les modes d'oxcillations $
\alpha^\mu_m$ et  $ {\tilde \alpha}^\mu_m$
 de $ X^\mu_L$,  $ X^\mu_R$ et   v{\'e}rifient  les relations de commutations
suivantes
\be
[{\tilde\a}^\mu_m,{\tilde
\a}^\nu_n]=[{\a^\mu}_m,\a^\nu_n]=m\delta_{m+n,0}\eta^{\mu\nu} \ee
\begin{equation}
[ \tilde \alpha^\mu_m,  \alpha^\nu_n]=0
\end{equation}
et
\begin{equation}
{ \a^\mu_n}^+=  \a^\mu_{-n},\quad {   \alpha^\mu_n}^+ =
{\alpha}^\mu_{-n}
\end{equation}
\begin{equation}
 {  {\tilde  \alpha}^\mu_n}+ = { \tilde \alpha}^\mu_{-n}.
\end{equation}
L'espace de Fock construit {\`a} partir de l'\'etat de vide
$|0>=|0,p>_L \otimes   |0,p>_R $  avec
\be
 \alpha_n|0>=\tilde \alpha_n|0>=0,\quad n>0,
\ee contient    des \'etats  de la forme:
\begin{equation}
  \prod \limits _{i=1}^k  \prod \limits _{j=1}^l \alpha_{-n_i}^{\mu_i}
\tilde \alpha_{-n_j}^{\mu_j}|0,k^\nu>.
\end{equation}
Comme dans le cas de la corde ouverte,  l'invariance   sous les
diff\'eomorphismes de la surface d'univers d'une corde ferm{\'e}e
exige que
 \bea
L_n|0>=\tilde L_n|0>=0,\quad n>1\\ L_0|0>=a|0>,\quad  \tilde
L_0|0>=a|0>. \eea Les $L_n$ et $\tilde L_n$ sont les modes de
Fourier des   composantes  gauches
  $T_L$ et droites  $T_R$ du tenseur \'energie-impulsion donn\'es par
\be
L_n={1\over2}\sum \limits_
{m=-\infty}^{+\infty}\alpha_{m-n}\alpha_n \ee
\be
L_0=-{\alpha' \over4} p^2+\sum \limits_ {n=1 }^{+\infty}
\alpha_{m-n}\alpha_n-a \ee
\be
{\tilde L}_n={1\over2}\sum \limits_
{m=-\infty}^{+\infty}\tilde\alpha_{m-n}\tilde\alpha_n \ee
\be
{\tilde L}_0=-{\alpha' \over4} p^2+\sum \limits_ {n=1 }^{+\infty}
\tilde \alpha_{m-n}\tilde \alpha_n- \tilde a, \ee o\`u $a=\tilde
a=1$.  La contrainte  $ L_0= {\tilde L}_0$  implique que la
formule de masse
  se r\'eduite  alors \`a
\begin{equation}
M^2={2\over\alpha'}(N_L+N_R-2),
\end{equation}
o{\`u} $N_L=\sum\alpha_{-n} \alpha_n$ et $N_R=\sum{\tilde
\alpha}_{-n} {\tilde \alpha}_n$ d\'esignent les nombres
d'oscillations gauches et droites respectivement.
 \subsection {Spectre des \'etats   }
 L'analyse de  l'{\'e}quation (1.2.20) pour la corde bosonique montre que:
\begin{itemize}
\item {L'{\'e}tat fondamental   $|0>$   $(N_L=N_R=0)$
de l'espace de Fock  correspond  {\`a} une  particule de masse
carr{\'e}e n{\'e}gative
 $M^2=-{4\over\alpha'}$. C'est une  particule scalaire appel\'ee
tachyon.}\item
 {L'\'etat excit\'e  $(N_L=1, N_R=0$ ou $ N_L=0, N_R=1)$,
d\'ecrivant
   une particule vectorielle,  a aussi une masse carr\'ee  n\'egative donc \`a
\'eliminer
    dans tout mod\`ele de corde consistante.}
 \item {L'{\'e}tat excit{\'e} $ (N_L=1, N_R=1)$ correspond  {\`a} des
particules de masse nulle
  s'{\'e}crit sous la forme
\begin{equation}
 \epsilon _{\mu \nu }\alpha_{-1}^\mu \tilde \alpha_{-1}^\nu|0,k^\nu>
\end{equation}
 dont le spin d\'epend du choix  du  tenseur de polarisation $\epsilon _{\mu
\nu }$ \cite{GSW,Pol1,Pol2,Kir}.
   Physiquement  ces \'etats non massifs sont class\'es dans des
repr\'esenations  du sous
     groupe transverse $ SO(24)$,  obtenu apr\`es  la  d\'ecomposition  du
groupe de  Lorentz
      $  SO(1,25)$
  $$
  SO(1,25)\to SO(24) \otimes SO(2)
$$
 en \'eliminant  les (ddl) non physique,  comme suit
\be
\a_{-1}^i{\tilde \a}_{-1}^j|0>= \a_{-1}^{[i}{\tilde
\a}_{-1}^{j]}|0>+\[\a_{-1}^{\{i}{\tilde \a}_{-1}^{j\}}-{1\o
24}\delta ^{ij}\a_{-1}^k{\tilde \a}_{-1}^k\]|0> +{1\o
24}\delta^{ij} \a_{-1}^i{\tilde \a}_{-1}^i|0>, \ee o\`u d'une fa\c
con \'equivalente
\begin{equation}
24_V \times 24_V=1\oplus 276\oplus 299.
\end{equation}
La partie sym{\'e}trique de trace nulle d\'ecrit le   graviton  $
g_{ij} $,
  le  tenseur antisym{\'e}trique d'ordre 2  est repr\'esent\'e par  le champ
    $ B_{ij} $ et  la   trace est interpr\'et\'ee  comme  le  dilaton   $
\phi$.}
\item{ Les {\'e}tats  \'excit\'es d'ordre sup\'erieur $ (N_L=N_R>1)$
ont tous  des masses
  tr\`es grandes puisque $ M\sim {1\over {\alpha}'^2}$.}
\end{itemize}

\chapter{  Mod\`eles de supercordes}
\pagestyle{myheadings}
\markboth{\underline{\centerline{\textit{\small{  Mod\`eles de
supercordes}}}}}{\underline{\centerline{\textit{\small{  Mod\`eles
de supercordes}}}}} \section {  Corde fermionique }
 La th{\'e}orie des cordes bosoniques
poss{\`e}de essentiellement deux probl{\`e}mes
 \cite{GSW,Pol1,Vafa1}:  L'existence de tachyon (particule non physique, qui
apparait tr{\`e}s naturellement
  dans  l'approche de  la quantification covariante de la th\' eorie de la
corde bosonique)   et
l'absence des particules fermioniques n\'ecessaires pour
d\'ecrire la mati\`ere. \\ La th{\'e}orie  des supercordres
r\'esoud  ces probl{\'e}mes en ajoutant des champs
  fermioniques  $  \psi^\mu(\tau,\sigma)$ sur la surface d'univers.   Cette
   augmentation par  des spineurs donne lieu \`a   une  sym\'etrie   plus
riche \`a savoir
     la sym\'etrie superconforme.   Notons  que  la supersym{\'e}trie demande
autant de
      d{\'e}gr{\'e}s de libert\'e fermioniques  que bosoniques.  De plus
l'adjonction  des champs fermioniques $ \psi^\mu(\tau,\sigma)$
{\`a} la  corde bosonique,
  qui sont {\`a} la  fois des spineurs de Majorana {\`a}  deux dimensions et
    des vecteurs de Lorentz,  impliquent  la r\'eduction  de  la dimension
critique
      de 26   {\`a}  $D=10$.   Ces champs fermioniques  peuvent {\^e}tre
p{\'e}riodiques
        ou antip{\'e}riodiques,  d\'efinissant  les secteurs de  Neveu-Schwarz
({\bf  NS})
         et de Ramond ({\bf R}) respectivement.   Pour construire une
th{\'e}orie consistante,
           nous introduisons de plus  la projection   de Gliozzi, Scherk et
Olive   {\bf (GSO)}
            qui {\'e}limine en particulier  le tachyon du spectre  et garantit
l'existence
             d'un nombre {\'e}gal de particules bosoniques et fermioniques \`a
chaque niveau
              d'excitation \cite{GSO1,GSO2}.
\subsection {  Corde fermionique classique}
Rappelons que dans la jauge conforme,  l'action de la corde
bosonique est donn\'ee par: $$ S= -{1\over2 \pi}\int d\tau d\sigma
\partial_{\alpha}X^\mu\partial^\alpha X_{\mu}.$$
 Cette action peut \^etre  g\'en\'eralis\'ee en introduisant les champs fermioniques
libres
  $  \psi^\mu(\tau,\sigma)$:
\begin{equation}
 S= -{1\over2 \pi}\int d \tau d\sigma ( \partial_{\alpha}X^\mu\partial^\alpha
X_{\mu}-i \bar\psi^\mu \rho^\alpha\partial_\alpha\psi_\mu),
\end{equation}
d\'ecrivanat  l'action de la supercorde classique.
  $\psi^\mu_A= {\psi^\mu_- \choose\psi^\mu_+}$ est un spineur de Majorana
    {\`a}  deux dimensions v{\'e}rifiant   l'\'equation de  Dirac
\begin{equation}
  \rho^\alpha\partial_\alpha \bar \psi^\mu_A=0,
\end{equation}
 o{\`u} $ \bar\psi =\psi^+ \rho^0$  et les  $\rho^\alpha$   satisfont
l'alg\`ebre de Clifford
\begin{equation}
 \{ \rho^\alpha,\rho^\beta\}=-2\eta^{\alpha\beta}.
\end{equation}
  Dans notre cas,  les  $\rho^\alpha$ ( $ \alpha=0,1$)   sont des matrices
$2\times2$
\begin{equation}
 \rho^0 =\left(\matrix{
0&-i\cr i&0\cr}\right), \quad \rho^1 =\left(\matrix{ 0&i\cr
i&0\cr}\right).
\end{equation}
 Les champs $X^\mu(\tau,\sigma)$ et $\psi^\mu(\tau,\sigma)$ sont des vecteurs
  de Lorentz dans l'espace-temps {\`a} $10$ dimensions.  En plus de la
sym{\'e}trie
   conforme, l'action   de la supercorde  (2.1.1) est invariante par les
transformations
     supersym{\'e}triques
\begin{equation}
\begin {array}{lcr}
\delta X^\mu =\bar\epsilon\psi^\mu\\ \delta\psi^\mu
=-i\rho^\alpha \partial_\alpha X^\mu\epsilon,
\end{array}
\end{equation}
 o\`u  $\epsilon$  est un spineur de Majorona.  Cette transformation  permet
de relier les bosons aux fermions.
La variation de l'action  de supercorde  conduit \`a
\begin{eqnarray}
\delta S= \frac {1}{\pi}\int d^2\sigma
(\partial_{\alpha}\bar{\epsilon})J^{\alpha},
\end{eqnarray}
o\`u $J^{\alpha}$ est le supercourant de Noether conserv\'e.   Ce
dernier est   donn\'e par
\begin{eqnarray}
J^{\alpha}=\frac
{1}{2}\rho^{\beta}\rho_{\alpha}\psi^{\mu}_{\alpha}\partial_{\beta}
X_{\mu},
\end{eqnarray}   qui  est un spineur \`a deux  composantes v\'erifiant
l'identit\'e:
$$\rho_{\a}J^{\alpha A}=0,$$ o\`u
$\rho_{\a}\rho^{\beta}\rho^{\a}=0$.  Alors que le  tenseur
\'energie impulsion s'ecrit sous la forme:
\begin{eqnarray}
T_{\alpha\beta}=\partial_{\alpha}X^{\mu}\partial_{\beta}X_{\mu}+\frac{i}{4}\bar{\psi}^{\mu}\rho_{\alpha}
\partial_{\beta}\psi_{\mu}+\frac{i}{4}\bar{\psi}^{\mu}\rho_{\beta}
\partial_{\alpha}\psi_{\mu}-\frac
{1}{2}\eta_{\alpha\beta}T^{\sigma} _{\sigma}.
\end{eqnarray}
L'action de la supercorde poss\`ede une sym\'etrie plus large que
l'alg\`ebre de Virasoro  de la corde bosonique. C'est l'alg\`ebre
superconforme engendr\'ee par les deux courants $T_{\alpha\beta}$
et $J_{\alpha}$.\\
\\
 {\bf Conditions aux bords et les  modes de
Fourier}\\
 Comme dans le cas des champs bosoniques $ X^\mu (\tau,\sigma)$, les champs
fermioniques
   doivent satisfaire des conditions aux bords. Ces conditions aux limites du
spineur de
    Majorana $\psi ^\mu (\tau,\sigma)$ sont obtenues en variant l'action de la
supercorde (2.1.1). Ceci donne lieu \`a  l'{\'e}quation  de Dirac et
\`a des termes de bords.
 \subsubsection{ ($\imath$)  Supercorde ouverte}
 Dans ce cas,    nous avons
\begin{equation}
 \delta S_F= \int d\tau [\psi^\mu_+\delta \psi_{\mu+}-\psi^\mu_-\delta
\psi_{\mu-}]^{\sigma=\pi}_{\sigma=0},
\end{equation}
  c'est  une  collection de quatre termes, puisque les deux extr{\'e}mit{\'e}s
de la supercorde sont ind{\'e}pendantes.
 Ainsi nous avons plusieurs  fa\c cons   d'annuler $\delta S_F$. Une fa\c con
de faire est de consid\'erer
\begin{equation}
 [\psi_+=\pm\psi_-]_{\sigma=0,\pi}
\end{equation}
et
\begin{equation}
[ \delta \psi_+=\pm \delta \psi_-]_{\sigma=0,\pi}.
\end{equation}
Si nous posons $ \psi_+(0,\tau)=\psi_-(0,\tau) $, nous aboutissons
{\`a} deux secteurs distincts:\\ Secteur  de {Ramond
({\bf R})} est d\'efinit par
 \begin{equation}
\begin {array}{lcr}
 \psi_+^\mu(\tau,0)=\psi_-^\mu(\tau,0)\\
\psi_+^\mu(\tau,\pi)=\psi_-^\mu(\tau,\pi).
\end{array}
\end{equation}
 Dans ce secteur, les oscillations fermioniques droite et gauche  sont
p\'eriodiques et   se d\'eveloppent en modes normaux sous la forme suivante
 \begin{equation}
\begin {array}{lcr}
 \psi_+^\mu(\tau,\sigma)={1\over
{\sqrt2}}\sum\limits_{n=-\infty}^{n=+\infty}d^\mu_ne^{-in(\tau+\sigma)}\\
 \psi_-^\mu(\tau,\sigma)={1\over
{\sqrt2}}\sum\limits_{n=-\infty}^{n=+\infty}d^\mu_ne^{-in(\tau-\sigma)},n\in Z.
\end{array}
\end{equation}
Secteur  de {Neveu-Schwarz ({\bf NS})} est d\'efinit par
\begin{equation}
\begin {array}{lcr}
 \psi_+^\mu(\tau,0)=\psi_-^\mu(\tau,0)\\
\psi_+^\mu(\tau,\pi)=-\psi_-^\mu(\tau,\pi).
\end{array}
\end{equation}
Dans ce secteur, les oscillations fermioniques droites et gauches,
sont antip\'eriodiques et  se d\'eveloppent   en modes normaux de
la fa\c con suivante
 \begin{equation}
\begin {array}{lcr}
 \psi_+^\mu(\tau,\sigma)={1\over
{\sqrt2}}\sum\limits_{r=-\infty}^{r=+\infty}b^\mu_re^{ir(\tau+\sigma)}\\
 \psi_-^\mu(\tau,\sigma)={1\over
{\sqrt2}}\sum\limits_{r=-\infty}^{r=+\infty}b^\mu_re^{ir(\tau-\sigma)},\quad
r\in Z+{1\over2}.
\end{array}
\end{equation}

\subsubsection{ ($\imath\imath$) Supercorde ferm\'ee}
 Dans le cas des supercordes ferm\'ees,
les conditions aux bords pour les fermions sont p\'eriodiques ou
antip\'eriodiques. Ceci conduit aux d\'eveloppement en modes
suivants\\ $\spadesuit$ \emph{fermions p\'eriodiques}
\begin{eqnarray}
\psi_{-}^{\mu}(\tau,\sigma)&=&\frac{1}{\sqrt{2}}\sum_{n\in
Z}d_{n}^{\mu}e^{-in(\tau-\sigma)}\nonumber\\
\psi_{+}^{\mu}(\tau,\sigma)&=&\frac{1}{\sqrt{2}}\sum_{n\in
Z}\tilde{d}_{n}^{\mu}e^{-in(\tau+\sigma)}
\end{eqnarray}
$\spadesuit$ \emph{fermions antip\'eriodiques}
\begin{eqnarray}
\psi_{-}^{\mu}(\tau,\sigma)&=&\frac{1}{\sqrt{2}}\sum_{r\in
Z+\frac{1}{2}}b_{r}^{\mu}e^{-ir(\tau-\sigma)}\nonumber\\
\psi_{+}^{\mu}(\tau,\sigma)&=&\frac{1}{\sqrt{2}}\sum_{r\in
Z+\frac{1}{2}}\tilde{b}_{r}^{\mu}e^{-ir(\tau+\sigma)}.
\end{eqnarray}
 Les conditions aux bords pour les modes \`a mouvement gauche et
ceux \`a mouvement droit peuvent \^etre choisies ind\'ependamment.
Par cons\'equent, l'espace de Fock total (secteur gauche
$\otimes$  secteur droit) de la supercorde ferm\'ee poss\`ede
quatre secteurs:
\begin{center}
\begin{tabular}{|c|c|}
  \hline
  \textrm{secteur} &   \\
  \hline
 {\bf  NS-NS} & $\{b_r^\mu, {\tilde  b}_s^\nu\}  \quad   r,s \in {\bf Z}+{1\o
2} $
  \\
  \hline
 {\bf  NS-R} & $\{b_r^\mu, {\tilde  d}_n^\nu\}  \quad   r \in {\bf Z}+{1\o 2},
n  \in {\bf Z}$
  \\
  \hline
{\bf  R-NS} & $\{d_m^\mu, {\tilde  b}_s^\nu\}  \quad   s \in {\bf
Z}+{1\o 2}, m  \in {\bf Z}$
  \\
  \hline
 {\bf R-R} & $\{d_m^\mu, {\tilde d }_n^\nu\}  \quad   m,n \in {\bf Z}$\\
  \hline
\end{tabular}
\end{center}
 \subsection{Supercordes h\'et\'erotiques classiques}
 Nous avons expos{\'e}  dans le pragraphe pr\'ecedent  des mod\`eles o\`u
les  deux
  secteurs d'oscillations gauche et droite sont essentiellement
ind{\'e}pendants,  et
    sont supersym\'etriques. Il se trouve qu'il est possible de combiner un
secteur
     d'oscillations gauche non supersym{\'e}trique  avec un secteur
d'oscillations droit
      supersym\'etrique.  Ce mod\`ele d\'efinit  la supercorde
h{\'e}t{\'e}rotique \cite{GHMR}.
        Le secteur droit  de cette  th{\'e}orie correspond {\`a} celui de la
th{\'e}orie
          de supercorde {\`a} $D=10$.  Les degr{\'e}s de libert{\'e}    sont
alors :\\
$\spadesuit$ \emph{secteur gauche} \bea
\{X_L^\mu(\tau+\sigma),\mu=0,1,\ldots,9\}\cup\{X^I_L=
X^I_L(\tau+\sigma)\},\quad I=10,\ldots,25. \eea $\spadesuit$
\emph{secteur droit}
\begin{equation}
\{X^\mu_R=X^\mu(\tau-\sigma),\psi_-^\mu(\tau-\sigma),\quad
\mu=0,1,\ldots,9 \}.
\end{equation}
 Notons  que  la partie  $\{X_L^\mu\}$ combin{\'e}e  \`a  $\{X_R^\mu\}$ donne
les dix composantes non compactes
de l'espace-temps de la th{\'e}orie des supercordes h{\'e}t{\'e}rotiques. Les
$16$ coordonn{\'e}es suppl{\'e}mentaires
 $\{X_L^I\}$  de la partie bosonique sont interpr{\'e}t{\'e}es comme les
coordonn{\'e}es compactes  d'un  tore $ T^{16}$
de dimension $16$.  Pour d{\'e}finir une th{\'e}orie de supercorde
supersym{\'e}trique, il  est plus commode de fermioniser
 les 16 champs suppl{\'e}mentaires   $\{X_L^I\}$   en  32 fermions droits
$\lambda_+^A, A=1,...,32$.  Dans  cette formulation,
l'action de la th\'eorie de   supercorde  h{\'e}t{\'e}rotique s'\'ecrit comme:
\begin{equation}
S={-{1\over{2\pi}}}\int  d\sigma^2 (
\partial_{\alpha}X^\mu\partial^\alpha X_{\mu}-2i \psi^\mu_-
\p_+\psi_{\mu-}-2i\lambda_+^A\p\lambda ^A_+),
\end{equation}
 et   \`a l'avantage d'exhiber clairement   le r{\^o}le des coordonn\'ees
internes. En effet selon la p{\'e}riodicit{\'e}
ou antip{\'e}riodicit{\'e} de chacun des $32$ spineurs de Majorana-Weyl
$\lambda ^A_+$, nous  distinguons     deux    th\'eories
 de supercordes h\'et\'erotiques  ayant des groupes de jauge dont les
r\`eseaux de racines des alg\`ebres sont auto-duaux. Il s'agit de:\\
\\
{\bf Supercorde h\'et\'erotique $SO(32)$}  \\
 Dans ce cas,  tous les  $\lambda ^A_+,A=1,\ldots,32$   sont p{\'e}riodiques
ou antip{\'e}riodiques
\begin{equation}
\lambda ^A_+(\pi +\sigma)=\pm \lambda ^A_+(\sigma),A=1,\ldots,32.
\end{equation}
 Les champs $\lambda ^A_+$ sont maintenant invariants sous les rotations du
groupe $ SO(32)$, nous  obtenons  alors une th{\'e}orie avec
 une sym{\'e}trie interne $SO(32)$  dite la th{\'e}orie de supercorde
h\'et\'erotique  dont  le groupe de jauge est  $SO(32)$.\\
\\
{\bf Supercorde h\'et\'erotique $ E_8 \times  E_8$}  \\
 Les  32 champs fermioniques   se scindent  en deux groupes  de 16 champs, sur
lesquels  nous imposons  les m{\^e}mes conditions  aux bords
\begin{equation}
\begin{array}{clr}
\lambda ^A_+(\pi+\sigma)=\lambda ^A_+(\sigma);\quad
A=1,\ldots,16\\ \lambda ^A_+(\pi+\sigma)=-\lambda
^A_+(\sigma);\quad A=17,\ldots,32.
\end{array}
\end{equation}
Les champs $\lambda ^A_+$  se transforment alors  lin\'eairement
suivant les repr{\'e}sentations $(\underline{16},1)$ ou
$(1,\underline{16})$ du groupe $SO(16)\times SO(16)$.   En
r{\'e}alit{\'e} la sym{\'e}trie interne est plut\^ot  le groupe $
E_8\times E_8$.  Cette th{\'e}orie est   appel\'ee le mod\`ele de
la supercorde h\'et\'erotique ayant   comme  groupe de jauge $
E_8\times E_8$.
 \subsection{Supercorde quantique  }
L'existence   des fermions ({\bf NS}) et ({\bf R})  conduit  \`a
un espace de Fock  plus riche pour  la th\'eorie  des supercordes.
 Pour  le secteur  ({\bf NS})  de  la  th{\'e}orie des supercordes ouvertes,
les relations  d'anticommutations des modes normaux des
composantes fermioniques se propageant {\`a} droite et \`a  gauche
sont donn\'ees  par\footnote{ Plus les relations de commutation
entre les $\alpha_n^ \mu$  pour la partie
 bosonique
$$ [\alpha^{\mu}_m,\alpha^{\nu}_n]=m\delta_{m+n,0}\eta^{\mu\nu}$$ }:
\begin{eqnarray}
\{b^{\mu}_r,b^{\nu}_s\}=\eta^{\mu\nu}\delta_{r+s}\nonumber\\
\{\tilde{b}^{\mu}_r,\tilde{b}^{\nu}_s\}=\eta^{\mu\nu}\delta_{r+s},
\end{eqnarray}
 o{\`u} $b^\mu_{-r}(b_{r}^\mu)$  et ${\tilde b}^\mu_{-r}({\tilde b}_{r}^\mu)$,  $ r>0$ sont
  des oscillations de cr\'eation (d'annihilation) gauche et droite
respectivement.\\
 Dans le secteur  {\bf NS},  les modes de Fourier $L_n$ du tenseur
\'energie-implusion  et     du supercourant  ont maintenant des
  contributions fermioniques et sont \'egaux
\begin{eqnarray}
L_m=\frac{1}{2}\sum_{n=-\infty}^{+\infty}\alpha_{-n}\alpha_{m+n}+
\frac{1}{2}\sum_{r=-\infty}^{+\infty} (r+m/2)b_{r}b_{m+r}.
\end{eqnarray}
\begin{eqnarray}
G_r=\sum_{n=-\infty}^{+\infty}\alpha_{-n}b_{n+r}.
\end{eqnarray}
L'\'etat fondamental de   la supercorde {\bf NS} est d\'efini par
\be
\alpha^{\mu}_n |0>_{{\bf NS}}=0, b^{\mu}_r |0>_{{\bf NS}}=0 \quad
n,r>0, \ee alors qu'un \'etat   g\'en\'erique   \`a mouvement
gauche s'\'ecrit comme suit
\begin{eqnarray}
|\psi>=\Pi _{i} (\alpha_{-n_{i}})^{\mu_i}\Pi_{j}
(b_{-r_{j}})^{\mu_j}|0>_{{\bf NS}}.
\end{eqnarray}
 La formule de masse g\'en\'eralisant  l'\'equation (1.2.8) est donn\'ee par
\begin{equation}
M^2=\sum\limits_{n=1}^\infty
\alpha_{-n}\alpha_m+\sum\limits_{r={1\over2}}^\infty r
b_{-r}b_r-{1\over2},
\end{equation}
 o\`u  $-{1\over2}$    est  la constante d'ordre normale du secteur  ({\bf
NS}).   Par   analogie avec
 l'op{\'e}rateur bosonique $\sum\limits_{n=1}^\infty \alpha_{-n}\alpha_m$,
l'op{\'e}rateur
$ \sum\limits_{r={1\over2}}^\infty r b_{-r}b_r$ compte  les {\'e}tats
excit{\'e}s \`a un facteur    $r$  pr{\'e}s. \\
   L'{\'e}tat fondamental $|0>_{{\bf NS}}$  dans ce  secteur  est encore un
tachyon avec  une masse carr{\'e}e \' egale
 \`a   $ -{1\over 2}$.  Le premier {\'e}tat excit{\'e} est  un vecteur de
l'espace-temps  non massif  obtenu par l'action
 de $b_{-{1\over 2}}^\mu$ sur le vide $ |0>_{{\bf NS}}$. Notons que le secteur
de Neveu-Schwarz se comporte comme un secteur bosonique.\\
 Dans le secteur de  Ramond ({\bf R}), les relations d'anticommutations
canoniques s'\'ecrivent comme:
\begin{eqnarray}
\{d^{\mu}_m,d^{\nu}_n\}=\eta^{\mu\nu}\delta_{m+n,0}\nonumber\\
\{\tilde{d}^{\mu}_m,\tilde{d}^{\nu}_n\}=\eta^{\mu\nu}\delta_{m+n,0}.
\end{eqnarray}
Les g\'en\'erateurs de Virasoro  et le supercourant dans ce secteur sont comme
suit
\begin{eqnarray}
L_m=\frac{1}{2}\sum_{n=-\infty}^{+\infty}\alpha_{-n}\alpha_{m+n}+
\frac{1}{2}\sum_{n=-\infty}^{+\infty} (n+1/2m)d_{r}d_{m+n},
\end{eqnarray}
\begin{eqnarray}
F_m=\sum_{n=-\infty}^{+\infty}\alpha_{-n}d_{m+n},
\end{eqnarray}
 alors que  l'op{\'e}rateur  de masse est:
\begin{equation}
M^2=\sum\limits_{n=1}^\infty
\alpha_{-n}\alpha_n+\sum\limits_{n=0}^\infty n d_{-n}b_n.
\end{equation}
 Dans ce cas, la constante d'ordre normale est nulle.  Ceci signifie que
l'{\'e}tat  fondamental
  $ |0>_{\bf  R}$  est  un fermion  non massif puisque  nous pouvons agir  sur
 lui plusieurs fois par  $d^\mu_0$ sans
 changer sa masse.  Signalons au passage que  les modes d'oscillations
$d^\mu_0$ satisfont  les relations d'anticommutations suivantes
\begin{equation}
\{d^\mu_0,d^\nu_0\}=\eta^{\mu\nu}.
\end{equation}
Ces op{\'e}rateurs  v\'erifient  les relations d'anticommutations
des matrices de Dirac
  \`a $D=10$;  ce  sont des matrices $32\times 32$.   Comme tout les \'etats
excit\'es sont
   obtenus en agissant sur l'\'etat fondamental par les op{\'e}rateurs de
cr{\'e}ation,
    ils sont de masse carr\'ee positive  et ob{\'e}issent tous \`a
l'{\'e}quation de Dirac.
     Par cons{\'e}quent, le secteur de Ramond est un secteur purement
fermionique.
\subsubsection{   Projection  {\bf GSO}}
 Nous avons discut\'e les premiers {\'e}tats du spectre des  deux secteurs
({\bf NS}) et ({\bf R}).
 Puisque l'{\'e}tat  fondamntal du secteur ({\bf NS}) est un tachyon, la
question qui se pose est comment l'{\'e}liminer.
 Il se trouve  que dans le secteur  ({\bf NS}), le nombre d'oscillations
fermioniques ($ \sum\limits_{r={1\over2}}^\infty r b_{-r}b_r$)
 peut \^etre pair  $(2N)$ ou impair $(2N+1)$. La projection  de
Gliozzi-Scherk et Olive  ( {\bf GSO})    dans  le secteur  {\bf  NS  } consiste
\`a ne retenir que les nombres  impair  \'eliminant du coup    le tachyon du
spectre \cite{GSO1,GSO2}. Le nouveau  {\'e}tat  fondamental
 doit \^etre   alors  un \'etat  excit\'e non massif $ b_{- {1\o 2}}|0>_{ {\bf
NS}}$.  En pratique,  la   projection
  {\bf GSO}  fixe  la chiralit\'e par le biais de   l'op\'erateur $ (-1)^F$
\be
(-1)^F_{ {\bf NS}}=-(-1)^ {\sum\limits_{r={1\o2}}^\infty
b_{-r}b_r}, \ee qui agit sur les champs de la supercorde comme
\bea (-1)^F_{ {\bf NS}} X^\mu &=& X^\mu\nn\\ (-1)^F_{ {\bf NS}}
\psi^\mu &=& - \psi^\mu.\nn \eea La projection   {\bf GSO}
consiste  alors  \`a garder  uniquement les {\'e}tats $| \psi>$
invariants par
\be
(-1)^F_{ {\bf NS}}| \psi>=| \psi>_{{\bf NS}}. \ee Dans le secteur
{\bf R}, la  projection  {\bf GSO}   agit sur  les \'etats $|
\psi>$ du spectre de Ramond comme suit
\be
(-1)^F_{ {\bf R}}| \psi>=| \psi>_{{\bf R}}, \ee o\`u
\be
(-1)^F_{{\bf R}}=\Gamma^{11}(-1)^ {\sum\limits_{n=1}^\infty
d_{-n}d_n}. \ee
 $\Gamma^{11}$  est l'op\'erateur de chiralit\'e {\`a}  dix dimensions,
donn{\'e} par
\be
\Gamma^{11}=\Gamma^0\ldots\Gamma^{9}, \ee
 satisfisant  les relations suivantes
\be
\{\Gamma^{11},\Gamma^{\mu}\}=0,\quad \mu=0,\ldots,9, \ee
\be
{(\Gamma^{11})}^2=1. \ee L'op\'erateur $(-1)^F_{ {\bf R}}$
v\'erifie le commutateur $$\[(-1)^F_{ {\bf R}}, d_n\]=0.$$   La
projection    {\bf GSO}  assure que  la th\'eorie
  de supercorde ouverte est un mod\`ele  supersym\'etrique de l'espace-temps
contenant
  la th\'eorie de superYang-Mills \`a dix dimensions  comme limite $ (\a'\to
0)$.
\section{Classification des supercordes}
 Nous pouvons obtenir imm\'ediatement  le spectre  de masse  des particules
pour une
supercorde ferm{\'e}e \`a partir  de celui de la supercorde ouverte. Il  est
essentiellement
 obtenu en prenant deux copies de l'espace d'Hilbert  d'une supercorde
ouverte  avec la formule de masse   donn\'ee par (2.1.28) et
(2.1.32).
 En appliquant la projection    {\bf GSO}  s{\'e}par{\'e}ment dans chacun des
secteurs gauche et droit,
  nous obtenons diff\'erents mod\`eles   selon le choix du signe relatif de la
projection  {\bf GSO} entre les
deux secteurs et   d\`ependent aussi de   la supersym\'etrie  de  la surface
d'univers.  En g\'en\'eral,
nous distinguons 5 mod\`eles de supercordes ayant une supersym\'etrie de
l'espace-temps:\\
1- Mod\`eles ayant une supersym\'etrie de l'espace-temps $N=2$.
Selon les valeurs propres  $\eta_L$  et $\eta_R$ de l'op\'erateur
{\bf GSO} dans le secteur R, on distingue:\\ (a) La th\'eorie type
IIA correspond    $\eta_L=-\eta_R=\pm 1$   ayant une
supersym\'etrie de l'espace-temps de type $(1,1)$. C'est un
mod\`ele non chiral.\\ (b) La th\'eorie type IIB correspond
$\eta_L=\eta_R=\pm 1$   ayant une supersym\'etrie de
l'espace-temps de type $(0,2)$ ou  $(2,0)$. C'est un mod\`ele
chiral.\\ 2- Mod\`eles ayant une supersym\'etrie de l'espace-temps
$N=1$:\\ (c) Supercorde  h\'et\'erotique $SO(32)$ ou $ E_8 \t
E_8$. Ce mod\`ele  a  une supersym\'etrie  h\'et\'erotique de  la
surface d'univers et une supersym\'etrie $ N=1$ de
l'espace-temps.\\ (d) La th\'eorie   type I    $SO(32)$ ayant une
supersym\'etrie  de  la surface d'univers et une supersym\'etrie $
N=1$ de l'espace-temps de type $(0,1)$.
\subsection{Supercorde  type IIA  et  IIB}
 Rappelons que dans la jauge du c\^one de lumi\`ere, l'\'etat fondamental du
secteur
 { (\bf NS)} correspond \`a un vecteur $8_V$  du  groupe de Lorentz $ SO(8)$
alors que
  l'\'etat fondamental du secteur  { \bf (R)} correspond \`a un spineur $8_S$
ou $8_{C}$.
   Le spectre total du supercorde type II est  obtenu par  le produit tensoriel
    des repr{\'e}sentations gauche et droite  ayant  $ 16 \times 16= 256$
{\'e}tats
     non massifs $ 8 \times 8= 64$  pour  chaque secteur
$$256=(8_V\oplus 8_S)\otimes (8_V\oplus 8_S).$$

\subsubsection{  Secteur bosonique de la th\'eorie   de type IIA (B)}
 Il existe deux  types:\\
{(\bf $\a$) Bosons  du secteur   NS- NS }\\
 Ces bosons sont obtenus  par le produit tensoriel suivant
\be
   8_V\otimes 8_V= 1\oplus35\oplus 28,
\ee
  qui contient  $64$ d{\'e}gr{\'e}s de libert{\'e}  r\'eparti en
un  dilaton $\phi$,  un  graviton $g_{\mu\nu}$    et un  tenseur
antisym{\'e}trique $B_{\mu\nu}$  respectivement.
 Notons que IIA et IIB ont les   m\^eme bosons  {\bf  NS-NS}\\
 {\bf  $(\beta$)  Bosons   R-R} \\
  Ces bosons d\'ependent  du choix de chiralit{\'e} relative des spineurs $S$
ou $C$  gauche et droit,
c'est {\`a} dire: \bea (-1)^F_{\bf R}| S>&=&| S> \nn\\ (-1)^F_{\bf
R}|C>&=&-| C>. \eea Dans le cas de la th{\'e}orie  IIA,  le
produit de deux spineurs de chiralit\'e oppos{\'e}e  $(S\otimes
C$) donne une th{\'e}orie non chirale dont les degr\'es de
libert\'e
\be
8_S \otimes 8_{C}=8_V+56, \ee
 d\'ecrivant  un  vecteur de jauge $ A^{\mu}$ et un tenseur   3-forme
$C^{\mu\nu\rho}$  antisym{\'e}trique.
 Dans le cas de la th{\'e}orie   de type  IIB, nous obtenons plut\^ot   une
th{\'e}orie chirale dont  les degr\'es de libert\'e
\be
8_S\otimes 8_ S=1\oplus 28 \oplus 35, \ee
 sont r\'epartis en   un  champ scalaire $\chi$ (axion),  une  2-forme   $
\tilde B_{\mu\nu} $ antisym{\'e}trique et une
 4-forme $D_{\mu\nu\rho\sigma}$   antisym{\'e}trique auto-duale.
\subsubsection{ Secteur fermionique de la th\'eorie   de type IIA(B)}
 Les deux secteurs fermioniques {\bf R- NS}  et  {\bf NS- R}   sont identiques
  et contiennent un fermion et un gravitino comme le montre   la
d{\'e}composition suivante
\be
8_V\otimes 8_S=8_C+56_C. \ee
 Ces  degr\'es de libert\'e constituent  les partenaires supersym{\'e}triques
des secteurs
  bosoniques {\bf NS-NS} et  {\bf R-R}.  Notons au passage que   les bosons
{\bf NS-NS} et
    {\bf R-R}
sont   tr{\'e}s diff\'erents; les premiers se couplent \`a
 la corde fondamentale alors que les seconds sont li\'es \`a
 des objets solitoniques connus sous le non de  D$p$-branes.

\subsection{Supercorde h\'et\'erotique SO(32)  et $E_8\times E_8$}
 Rappelons  que  les d{\'e}gr{\'e}s de libert{\'e} du secteur d'oscillations
gauche sont
  constitu{\'e}s par  celui de la corde bosonique \`a  $D=26$ dimensions,
alors que celui
    du secteur droit est form{\'e} par celui de  la supercorde \`a dix
dimensions. Les deux
     secteurs ne  sont coupl\'es que   par  les modes  z{\'e}ro des
oscillations bosoniques
      qui donnent naissance aux impulsions $(P_L,P_R)$. Evidemment
l'\'egalit\'e $(P_L=P_R)$
        n'est plus verifi\'ee du fait que  le $P_L$ est \`a  $ 26 $
dimensions, alors que
         $(P_R)$ est \`a dix dimensions.  Il est  alors  commode   de
d{\'e}composer $(P_L)$
 en deux blocs de composantes $10+16$   comme suit
\bea
 SO(1,25)\to& SO(1,9)\otimes SO(16),\nn\\
{\underline {26} }\to& ({\underline {10}},{\underline
{1}})\oplus({\underline {1}},{\underline
 {16}}),\\
P_L^{26}\to& P^{10}_L\oplus P^{16}_R,\nn \eea
 o\`u $ P_L^{10}$  est  identifi\'ee avec $ P_R$.  Les 16  d{\'e}gr{\'e}es de
libert{\'e}
  $ P^{16}_L$  suppl{\'e}mentaires sont  interpr\'et\'es comme les
g\'en\'erateurs  d'une
   alg\`ebre torique maximale d'une alg\`ebre de Lie. L'invariance modulaire
restreint
     cet alg{\'e}bre \`a   $ SO(32)$  ou $ E_8\times E_8$.  Dans la jauge de
c\^one de  lumi\`ere, les  conditions de masses s'ecrivent \bea M^2 ={1 \o
2}P^2_A+N_L-1=N_R-a_R, \eea o{\`u} $N_L$ et $N_R$ sont  les modes
d'oscillations   gauche et droit,  $P_A$  est l'impulsion  interne
et $a_{  R}$ est la constante d'ordre normal ($a_{ R}= {1 \o 2}$
pour ({\bf  NS}) et $a_R=0$ pour ({\bf R})). Les \'etats non
massifs dans  le secteur de propagation droite de la th{\'e}orie
des supercordes h{\'e}t{\'e}rotiques sont comme pour la supercorde
de type II \`a  savoir:
 \be
{\underline V}\oplus {\underline S}, \ee
 o{\`u} ${\underline V}={\underline {10}} $ est un vecteur de $ SO(1, 9)$  et
{\underline S}  est un spineur de
 $ SO(10)$ de chiralit{\'e} bien d\'efinie:
\bea {\underline S}=&{\underline S_+}\oplus {\underline S_-}\nn\\
{\underline {16}}=&{\underline 8_+}\oplus {\underline 8_-}. \eea
 Alors que les \'etats de masse nulle  du  secteur gauche proviennent de deux
possibilit\'es:\\
1- $ N_L=1 $ et $ P^2_A=-m^2=0$:\\
 Dans cette situation, les {\'e}tats ${\underline {16}}\oplus {\underline
{V}}$, correspondent aux champs de
 jauge du groupe de jauge  $ U(1)^{16}$.  Ces \'etats sont donn\'es par
$\a_{-1}^A|0>$. \\
2- $ N_L=0 $ et  $P^2_A=2$:\\
  Ce cas correspond aux poids non nuls de la repr\'esentation adjointe du
groupe
   G ($G= E_8 \t E_8$ ou $ SO(32))$.
    En combinant  les deux cas $ N_L=1 $ et  $P^2_A=2$,  nous obtenons
     $ {\underline {480}}+ {\underline{16}}={\underline {496}}$ {\'e}tats se
     transformant suivant la repr{\'e}sentation adjointe du groupe G ($ E_8 \t
E_8$ ou
      $ SO(32)$). Ceci  conduit aux \'etats
\be
{\underline {V}}\oplus {\underline { adj G}}. \ee Le produit
tensoriel  des propagations  gauche et droite  donnera les
\'etats non
 massifs suivants
\be
({\underline {V}}\oplus {\underline { adj G}})\otimes ({\underline
V}\oplus {\underline S}). \ee
 Les \'etats bosoniques,   obtenus par le  produit tensoriel $({\underline
{V}}\oplus {\underline { adj G}})\otimes {\underline V}$,
sont donn\'es par
 \be
( g_{\mu\nu}, B_{\mu\nu},\phi)+A_\mu  (A_\mu=A_\mu^a T_a,\quad a=1,\ldots
,\mbox {dim de }SO(32) (ou \; E_8 \t E_8 )). \ee
 Alors  que la partie fermionique est obtenue  par le produit  $({\underline
{V}}\oplus {\underline { adj G}})\otimes {\underline S}$,
qui fournit le gravitino et le jaugino, partenaires supersym{\'e}triques du
graviton et du champ de jauge respectivement, dans la
repr\'esentation adjointe du groupe de jauge. Gr\^ace \`a la d\'ecouverte de
la th\'eorie des supercordes h\'et\'erotiques, il est
 devenu possible de construire des mod\`eles ph\'enom\'enologiques
reproduisant les trois g\'en\'erations de mati\`ere chirale.
  \subsection {Supercorde  ouverte type I}
 Nous avons jusqu'{\`a} pr{\'e}sent  d{\'e}crit le spectre des {\'e}tats  non
massifs de
  la  th{\'e}orie des supercordes ferm\'ees.  Nous compl\'etons cette \'etude
 par
   la description du sepctre  de la supercorde type I. Ce spectre  est  obtenu
 {\`a}
    partir des supercordes ferm{\'e}es  de type IIB.   Pour cela,
consid{\'e}rons
     l'op{\'e}rateur $\Omega$ qui change la direction de la corde,   ainsi
que  les modes
      d'oscillations droite et gauche
\bea \Omega: \quad  \sigma \to 2\pi-\sigma,\quad \Omega^2=1, \eea
 o\`u d'une  fa\c con  \'equivalente
\be
 X_L^\mu \to X_R^\mu,
\ee
\be
\Omega \a_{n}^\mu\Omega ^{-1}= {\tilde \a}_{n}^\mu. \ee
 Nous allons garder uniquement les {\'e}tats $|\Phi>$  de l'espace d'Hilbert
qui sont invariants  par $\Omega$:
\be
\Omega|\Phi>=|\Phi>. \ee
Ces \'etats  repr{\'e}sentent   la projection  de l'espace d'Hilbert sur le
sous espace invaraint par l'op\'erateur de projectioin $ P= {1\over2}(1+\Omega)$.   En
pratique,
   ceci   revient {\`a} prendre le quotient par involution $z \to\bar z$, avec
$z=\tau+i\sigma$,
     $z=\tau-i\sigma$,  inversant l'orientation de la surface d'univers.  Dans
le cas  des
        th{\'e}ories de type II, cette op{\'e}ration doit {\^e}tre combin\'ee
avec une
         involution $ (-1)^F$ sur l'espace de Fock des {\'e}tats fermioniques
en {\'e}changeant
         les fermions gauches et droites.  Ce qui n'est  possible   qu'en
th\'eorie type IIB   o\`u les fermions ont la m{\^e}me
chiralit\'e.   Notons au passage que cette projection {\'e}limine
la moiti{\'e} des  charges supersym\'etriques de la th{\'e}orie de
type IIB conduisant ainsi \`a  une  th\'eorie du type I.
 Nous allons \'etudier seulement le spectre  bosonique de la  th\'eorie de type
  IIB qui reste apr\`es projection.\\
 Dans le secteur {\bf  NS-NS},  nous avons
\be
 \Omega \a_{n}^\mu {\tilde \a}_{n}^\nu\Omega ^{-1}={\tilde
\a}_{n}^\nu\a_{n}^\mu,
\ee
 ceci  montre que  seulement la partie sym{\'e}trique du produit tensoriel
{\bf NS}-{\bf NS}
     qui est invariante  sous  $\Omega$
\be
{\bf NS-NS}\longrightarrow (8_V\otimes 8_V)_{sym}=(\phi,g_{\mu\nu}). \ee
Dans le secteur {\bf  R-R},  comme les d{\'e}gr{\'e}s de
libert{\'e} sont fermioniques,
  nous obtenons   dans  chaque partie (gauche,  droite ) un signe (-)
suppl{\'e}mentaire
 {\'e}changeant les {\'e}tats d'oscillations  gauches et droites. Ceci
signifie que  nous devons prendre
uniquement la partie antisym{\'e}trique du produit tensoriel
\be
{\bf R- R}\longrightarrow (8_V\otimes
8_V)_{antisym}=(\tilde B_{\mu\nu}). \ee
 Dans le but  de construire une th\'eorie  consistante,  nous devons
introduire un secteur contenant des
 supercordes ouvertes.
 Rappelons que dans le cas de la th\'eorie de corde ouverte,  l'op\'erateur
$\Omega$ est donn\'e par
\bea \Omega: \quad \sigma \to \pi-\sigma, \eea
 ainsi le champ bosonique   de la forme suivante
$$ X^{\mu}=\sum\limits_n {\a}_n^{\mu} cosn\sigma .$$  Par
cons\'equent,  l'action $\Omega$ sur les oscillations
$\a_n^{\mu}$ peut {\^e}tre  \'ecrite comme suit
\be
\Omega \a_{n}^\mu\Omega ^{-1}= (-1)^n { \a}_{n}^\mu. \ee Nous
pouvons facilement  voir  que l'\'etat non massif   $ \a
_{-1}^\mu|0,k>$, qui correspond aux champs de jauge,   n'est pas
invariant par $\Omega$ \`a cause d'un signe (-). Ce probl\`eme
peut \^etre surmont\'e  gr\^ace   \`a  l'existence  des bords
permettant le couplage \`a un champ de jauge par l'interm\'edaire
des
 charges ponctuelles dites de {\it  Chan-Paton }. Dans le cas des cordes
ouvertes
  non orient\'ees,  ceci est  en accord avec l'existence d'un champ de jauge
dans
   le spectre des \'etats de  masses nulles.  Dans ce cas,  le secteur de
jauge  fournit un
     champ   $ A_{\mu}^{ij}$ avec  deux indices suppl\'ementaires  provenant
des extremit\'es
      $(i,j)$ de la supercorde ouverte:
\be
 A^{\mu}_{ij}=\a_{-1}^\mu |0,k>\t |ij>.
\ee Encore la projection $\Omega$ transforme cet \'etat \`a un
signe (-)  pr\`es  mais en changeant  l'indice $i $ en  $j$
\be
\Omega \a_{-1}^\mu |0,k,i,j>\Omega^{-1}=-\a_{-1}^\mu |0,k,j,i>, \ee
 o\`u d'une fa\c con \'equivalente
\be
A_{\mu}^{ij}=-A_{\mu}^{ji}, \ee et par cons\'equant  la partie
antisym\'etrique  qui  est  invarainte.
   Le  crit\`ere  d'\'elimination   des anomalies  dans  la  th\'eorie
     de supergravit\'e de type I restreint le groupe de jauge \`a $SO(32)$.
Nous obtenons
       alors la th{\'e}orie des supercordes de type I  dont le spectre de
masse nulle
        correspond  au dilaton $\phi$,  graviton $ g_{\mu\nu}$  du secteur de
{\bf NS-NS} de
         la th{\'e}orie de type IIB,  le  tenseur antisym{\'e}trique  $\tilde
B_{\mu\nu} $
           du secteur  {\bf  R-R} des supercordes ferm\'ees, les champs  de
jauge $SO(32)$ du secteur
 des cordes ouvertes et leurs partenaires fermioniques sous la
supersym{\'e}trie $N=1$ \`a dix dimensions.   La partie bosonique  de ce
mod\`ele est donn\'ee par
\begin{equation}
(g_{\mu\nu},\tilde B_{\mu\nu},\phi)+A_\mu (A_\mu=A_\mu^a T_a,\quad
a=1,\ldots  ,\mbox {dim de }SO(32)).
\end{equation}
Finalement, nous signalons que  la th{\'e}orie de supercorde de
type I
  a un spectre identique \`a celui  de la th{\'e}orie de supercorde
h\'et\'erotique de type
    $ SO(32)$.

\chapter{Compactification des Mod\`eles de    Supercordes}
\pagestyle{myheadings}
\markboth{\underline{\centerline{\textit{\small{Compactification
des Mod\`eles de
Supercordes}}}}}{\underline{\centerline{\textit{\small{Compactification
des   Mod\`eles de     Supercordes}}}}}
 A ce stade, nous avons vu qu'il existe   cinq mod\`eles de supercordes
consistants:
   les deux model\`es de  supercordes ferm{\'e}es IIA et IIB,  le mod\`ele de
supercorde
    de type I  ayant un groupe de jauge $SO(32)$, contenant les cordes
ouvertes et ferm\'ees,
 et les deux  supercordes ferm{\'e}es h{\'e}t{\'e}rotiques
      $E_8 \times E_8$ et $SO(32)$.  Ces cinq  mod\`eles  de supercordes
vivent cependant  dans un
 espace-temps \`a dix dimensions non compact. Pour ramener  ces th\'eories au
monde r\'eel, nous  avons besoin de
les d\'efinir  dans notre espace habituel \`a 1+3  dimensions. De ce fait, il
faudrait  compactifier les six coordonn\'ees
 d'espace suppl\'ementaires.  Ainsi nous devons consid{\'e}rer des
g{\'e}om{\'e}tries  o\`u l'espace  de Minkowski
 {\`a} dix dimensions  $M_{10}$   se d{\'e}compose en une vari{\'e}t{\'e} non
compacte correspondante  {\`a} l'espace-temps
de Minkowski usuel $M_{4}$ et une vari{\'e}t{\'e} compacte $\tilde X_6$ de
dimension  $6$, et de volume tr{\`e}s petit devant
 notre {\'e}chelle d'observation,
$$ M_{1,9}\to M_{1,3}\times \tilde X_6. $$ C'est le sc{\'e}nario  de
compactification des supercordes se propageant dans
  $M_{1,9}$ {\`a} des supercordes se propageant  dans des espaces de type
$M_{1,3}\times \tilde X_6$.
Comme nous le verrons plus tard, nous allons au del\`a de ce sc\'enario en
consid\'erant  des compactifications non
  seulement \`a quatre dimensions mais \`a des espaces-temps arbitraires
$M_{1,9}\to M_{1, 9-d}\times \tilde X_d$.
Ces espaces compacts  $  \tilde X_d$  sont choisis de sorte que
les supercordes
 se  propagent dans un espace-temps de dimension  $(10-d)$  pr\'eservant un
certain
  nombre de supersym\'etrie d'espace-temps. Puisque un spineur \`a dix
dimensions se
   d\'ecompose en un spineur sur l'espace $\tilde X_d$ et un spineur \`a
$(10-d)$, il en r\'esulte que
le nombre de supersym\'etrie pr\'eserv\'e \`a   $(10-d)$ dimensions de
l'espace-temps  depond du groupe d'holonomie de
 $\tilde X_d$.  Cette propri\'et\'e  permet de classer les diff\'erentes
vari\'et\'es $\tilde X_d$  qui conduisent \`a des mod\`eles de
supercordes
\`a $(10-d)$ dimensions \cite{GSW,Pol1,Vafa1}.\\
Dans ce chapitre,  nous \'etudions   trois  exemples de
compactification  des supercordes     pr{\'e}servant  16 ou  8
charges  supersym\'etriques  parmi les 32 supercharges originales.
Il s'agit de:
\begin{itemize}
  \item {La compactification toroidale sur le tore  $T^d$ ayant un   groupe
d'holonomie trivial.  Le mod\`ele  r\'esultant aura alors 32
supercharges}.
  \item {  La compactification sur l'hypersurface $\tilde X_4$=K3 de groupe
d'holonomie $SU(2)$}.
  \item { La compactifiaction sur des vari{\'e}t{\'e}s de Calabi-Yau {\`a} six
dimensions (trois complexes) $\tilde X_6=Y_3$ ayant un groupe d'holonomie $SU(3)$}.
\end{itemize}
 Par ailleurs, puisqu'il existe cinq types de mod\`eles de supercordes
consistants
  \`a 10 dimensions et diff\'erent choix possibles de vari\'et\'es compactes $
\tilde X_d$,
   nous allons nous  trouver avec un grand nombre de mod\`eles de supercordes
de divers
    dimensions qui ont un nombre  de supersym\'etrie d'espace-temps
diff\'erent et un nombre
     de degr\'es de libert\'e aux  faibles \'energies diff\'erents.  Les
\'etats du vide de
       ces mod\`eles de supercordes  de dimension inf\'erieures  sont
diff\'erents de ceux
         \`a dix dimensions.  Ces vides sont d\'etermin\'es par les valeurs
moyennes des
          champs scalaires de masse  nulle dits champs de module.   L'ensemble
  de ces scalaires constitue   une  vari\'et\'e
 souvent  d\'enomm\'ee  espace des modules (moduli space).\\
La compactification offre  les possibilit\'es de:
\begin{itemize}
  \item {  R\'eduire
le nombre de charges
  supersym\'etriques.  } \item{    D\'ecrire  les mod\`eles de supercordes
\`a  divers  dimensions.}
    \item{ Donner naissance \`a des connections entre diff\'erents mod\`eles
de supercordes.}
\end{itemize}
   Cette derni\`ere propri\'et\'e  sera consid\'er\'ee  lorsqu'on
\'etudiera les   sym\'etries  de dualit\'es dans le chapitre cinq. Ces
derni\`eres  permettent
 de r\'esoudre  partiellement  le probl\`eme  de   l'absence  de la sym\'etrie
de jauge non ab\'elienne
en   th\'eorie perturbative  des  supercordes de type II.    Ainsi
il   indique  l'existence de
 l'aspect  non perturbatif des mod\`eles de supercordes  bas\'e sur l'\'etude
des objets  \'etendus appel\'es  {\it branes}
  dont les  masses  augmentent  proportionnellement    avec $ 1 \o g^2$, et
ils \'echappent donc au spectre perturbatif.
Le cas le plus int\'eressant de ces compactifications  est
observ\'e \`a six dimensions
   de la supercorde  h\'et\'erotique sur le tore $ T^4$ ou    la  supercorde
    de type IIA sur la surface  K3. Ces deux mod\`eles,   de supersym\'etrie $N=2$  \`a
six dimensions, poss\`edent
le m\^eme espace des modules. Cette  propri\'et\'e  est confirm\'ee   par
l'\'etude des solutions solitoniques
qui fournissent  le spectre  des \'etats  non perturbatifs  (D$
p$-branes)\cite{Vafa1}.
 \section{ Compactification toroidale des supercordes }
 Afin de mieux comprendre la compactification toroidale, nous commen\c cons
par le cas  simple du cercle $S^1$ de rayon $R$.
\subsection{Compactification  sur $ S^1 $}
 Dans ce cas,  la coordonn\'ee compacte de la corde est  donn\'ee  par
\be
X(\sigma+2\pi,\tau)=X(\sigma,\tau)+2\pi mR, \ee o\`u $ m \in  {\bf
Z}$. Apr\`es la  compactification sur  $S^1$,   le spectre comporte
  des \'etats de  masse $ M={m\o R}$   correspondant aux excitations des
champs portant
   un moment interne $   P={m\o R}$ suivant  la direction compacte. Il
contient \'egalement
     des \'etats d'enroulement de masse $ M={n R \o \a'}$ $ (n \in \bf  Z)$
correspondant
       \`a une supercorde enroul\'ee  $n$ fois autour du cercle $ S^1$.
         Dans la construction de l'espace de Hilbert     pour       $ S^1$
nous devons rel\^acher la condition $ P_L=P_R$.
Cette \'egalit\'e n'est plus valable du fait que la supercorde  ferm\'ee  $
X(\tau,\sigma)$ peut tourner autour du cercle $S^1$  sans
 revenir n{\'e}cessairement \`a sa position initiale. Les moments $( P_L,P_R)$
sont donn\'es  par
\be
( P_L,P_R)=({m\o {2R}}+nR,{m\o 2R}-nR). \ee
  Notons que nous avons  les  propri{\'e}t{\'e}s  typiques  suivantes:\\
$$ P_L \neq P_R,$$ $$ P_d^2-P_g^2=2mn \in 2{\bf  Z},$$
contrairement au cas des vari\'et\'es non compactes.\\
 Le spectre de toutes les valeurs  permises de $ ( P_L,P_R)$ est  invariant
par la  transformation $ R \to {1\o 2R}$ et  $ n\to m$.
 Cette sym{\'e}trie  est une   cons\'equence  de ce qu'on appelle   la
dualit\'e-T.  Elle  admet une extension remarquable dans le cas
 des compactifications toroidales sur un tore de dimension sup\'erieure.  En
effet, \'etant donn\'e  un tore $T^d=(S^1)^{\otimes d}$ de
 dimension $d$, tous les autres tores sont alors obtenus en faisant une
transformation de boost du groupe $ SO(d,d)$ sur les vecteurs
$( P_L,P_R)$.  Puisque   la rotation du  couple  $ ( P_L,P_R)$ par une
transformation de $ O(d) \t O(d)$ ne change pas le spectre des \'etats de
la supercorde et que  les transformations de Boost  $O(d,d,{\bf Z})$  ne
changent  pas le r\'eseau, il en d\'ecoule que l'espace des choix  des
  tores $T^d$ in\'equivalents   est  un espace homog\`ene de dimension $d^2$
donn\'e par
\be
{SO(d,d)\o SO(d)\t SO(d)\t SO(d,d,{\bf Z})}. \ee Le groupe
$SO(d,d,{\bf Z})$  g{\'e}n{\'e}ralise la dualit\'e-T
consid\'er\'ee dans le cas de  $S^1$  pour une   compactification
sur $T^d$.  Plus g{\'e}n{\'e}ralement,  la   dualit{\'e}-T
s'{\'e}tend aux compactifications sur des espaces de Calabi-Yau:
c'est  la sym\'etrie miroir.  Cette  derni\`ere   d'inter\^et
primordial  dans la nouvelle \'etude des   mod\`eles  de
supercordes    sera  trait\'ee   dans ce chapitre.   Dans ce  paragraphe
nous allons pr\'eciser comment la dualit\'e-T se manif\'este sur
les champs de Fermi.
 \subsection{Compactification des  mod\`eles
$N=2$ sur $T^d$}
  Dans cette  \'etude    nous allons voir que la compatification des mod\`eles
IIA et IIB sur un
 cercle de rayon $R$ ont  les m\^emes degr\'es de libert\'e aux faibles
\'energies. En fait, IIA et
IIB compactifi\'ees sur $S^1$ sont \'equivalentes puisqu'elles sont
interchang\'ees par la dualit\'e-T
 qui agit comme R  en  ${1 \o R}$. Cette sym\'etrie  agit aussi sur les
fermions le long de la direction compacte $S^1$:
\bea \psi _L^9&\to & -\psi_L^9\nn\\ \psi _R^9&\to &+ \psi_R^9.
\eea
 Cette \'equation montre que le produit des fermions gauches  $\psi _L^o\psi
_L^1\ldots \psi _L^9$,
d\'eterminant  la chiralit\'e d'espace-temps, se transforme en son oppos\'e.
Ceci  implique que sous
la projection  {\bf GSO},  nous  aboutissons  \`a  une  chiralit\'e oppos\'ee
\`a celle qu'on avait
auparavant  et    ainsi les spineurs  $ 8_S$ et  $8_C$ du secteur gauche
s'interchangent.  \\
\par Dans la compactification  sur $T^d$, les \'el\'ements du groupe de
dualit\'e $O(d,d, {\bf Z})$  qui
 changent  les mod\`eles IIA et IIB sont ceux  qui n'appartiennent pas au
groupe  $SO(d,d,{\bf Z})$.
Cela signifie que le groupe  $SO(d,d,{\bf Z})$  est  une sym\'etrie    de IIA
et  de IIB  s\'epar\'ement.
 Par contre, les \'el\'ements discrets de $O(d,d,\bf  Z)$ ne le sont pas.\\
Dans  la  compactification toroidale des   mod\`eles de
supercordes de type II,  les $d^2$ scalaires param\`etrisent
l'espace quotient  $ SO(d,d)\o SO(d)\t SO(d)$ correspondant  aux
choix de la m{\'e}trique $g_{ij}$,  ($d(d+1)\o2$ d{\'e}gr{\'e}s de
libert{\'e}), et celui du champ antisym{\'e}trique $ B_{ij}$, ( $
d(d-1)\o2$ d{\'e}gr{\'e} de libert{\'e}s), sur le tore $T^d$
respectivement.  Notons que le dilaton ainsi  les valeurs moyennes des tenseurs
antisym{\'e}triques de jauge du secteur de Ramond   fournissent
aussi des modules suppl{\'e}mentaires de la th{\'e}orie compactifi{\'e}e sur laquelle  la dualit{\'e}-T doit encore agir.\\
\\
{\bf Exemple 1:} IIA/ $T^4$\\
\\
A titre d'exemple, \'etudions  la compactification  du mod\`ele
IIA sur $T^4$.  Nous avons 16 param{\`e}tres, sp{\'e}cifiant
l'espace  ${SO(4,4,R) \o {SO(4) \t SO(4) }}$,   param\`etrisant
respectivement  les   valeurs moyennes  des  champs  $ g_{ij}$ et
$B_{ij}$ sur le tore $ T^4$.   En plus de ces param\`etres
provenant du secteur  {\bf  NS-NS}, nous avons aussi  8  (
$C^1_4+C^3_4=4+4$)  param{\`e}tres  sp\'ecifiant    le choix   des
champs de Ramond  $ A_{\mu}$  et  $C_{\mu\nu\lambda}$
respectivement.  Il faut encore  ajouter un scalaire qui
correspond  au  dilaton donnant lieu  \`a  25 scalaires
d{\'e}finissant l'espace totale  des   modules   IIA sur $T^4$
\be
{SO(5,5)\o SO(5)\t  SO(5)}. \ee
 Notons  au passage que cet espace des modules peut  \^etre aussi obtenu \`a
partir de la
 compactification de la th\'eorie de  supergravit\'e  $N=1$  {\`a}  onze
dimensions sur  le   tore $T^5$ \cite{Vafa1}.
\subsection{Compactification des mod\`eles   $N=1$ sur
$T^d$}
 Rappelons  qu'\`a dix  dimensions, nous avons trois  mod\`eles  de
supercordes ayant une
 supersym\'etrie $N=1$. Ce sont les  mod\`eles  h\'et\'erotiques  $E_8\t E_8$
et $SO(32)$ et type I $SO(32)$.
 Par exemple, dans le cas  de la supercorde h\'et\'erotique,  les bosons de
jauge \`a  10
 dimensions fournissent  apr\`es  compactification toroidale sur $T^d$
($16\times d$)
 modules associ\'es  aux  lignes de Wilsons brisant la sym{\'e}trie de jauge
de rang
16 par le m\'ecanisme  de Hosotani.   Nous devons ajouter  \`a ces modules
le choix de
 la m{\'e}trique $g_{ij}$, le dilaton  ainsi que le champ antisym{\'e}trique
$B_{ij}$
sur $T^d$ dont les modules forment  le m{\^e}me espace  quotient que celui du
secteur
{\bf NS-NS} de la th{\'e}orie de type II discut\'e auparavent  ($ SO(d,d)\o
SO(d)\t  SO(d) $).
 La structure  de l'espace des modules  total  est donn\'ee par
\be
{SO(d+16,d)\o SO(d)\t  SO(d+16)} \times SO(1,1), \ee modulo  le groupe de  la
dualit{\'e}-T {\'e}tendu $O(d+16,d, {\bf  Z})$.
  La construction de l'espace de Hilbert dans ce cas ne diff\`ere de celle de
type II {\`a} D=10 que par
le fait que $(P_L,P_R)$ appartient maintenant  {\`a} un r{\'e}seau  de Lorentz
 pair et auto-dual de signature
  $(d+16,d)$ appel{\'e} r{\'e}seau de Narain $\Gamma^{d+16,d}$ \cite{NSW}.
Notons aussi que lorsque le r\'eseau   $\Gamma^{d+16,d}$ est factoris\'e
en $\Gamma^{d,d}\oplus \Gamma^{16}$,
 pour des lignes de Wilson nulles, nous retrouvons la sym\'etrie de jauge
$SO(32)$ (ou $E_8 \t  E_8$)  de
la th\'eorie  \`a dix dimensions.
Il d\'ecoule de l'unicit\'e du  r\'eseau de  Narain  $\Gamma^{d+16,d}$  dans
la compactification toroidale que les th\'eories h\'et\'erotiques $ SO(32)$
 ou  $E_8 \t  E_8$ sont isomorphes  par  la  dualit\'e-T.\\
\\{\bf Exemple 2:  H\'et\'erotique  sur  $T^6$}\\
\\
A titre d'exemple, consid\'erons la compactification  sur $T^6$ donnant une
th\'eorie $N=4$  \`a  $D=4$ .
 Nous avons  $ 6\t 6=36 $ param\`etres sp\'ecifiant la m\'etrique $g_{ij}$, le
champs  $B_{ij}$ et le dilaton $\phi$.
En plus de ces  param\`etres, il faut ajouter $ 16\t 6=96$ param\`etres
provenant  des  lignes de Wilsons du champ de
jauge du groupe $ SO(32) $ ou $E_8\t E_8$.  L'espace des modules de la
supercorde h\'et\'erotique sur $T^6$ est donn\'e par
 \begin{equation}
 {SO(22,6, {\bf R})\over {SO(22,{\bf R}) \times SO(6,{\bf R})}} \times
{Sl(2,{\bf R})\o  SO(2,{\bf R})}.
\end{equation}
o\`u ${Sl(2,{\bf R})\o  SO(2,{\bf R})}$  correspond au  champ
complexe dilaton-axion.
 Finalement notons que pour la th\'eorie de type I, la partie locale de son
espace des  modules sur  $T^d$ est vue comme
 celle de la supercorde h\'et\'erotique et   admet  le m\^eme groupe de la
dualit\'e-T  $ SO(d+16,d,{\bf Z})$.
 \section{  Vari{\'e}t{\'e}s de Calabi-Yau }
La compactification toroidale que nous avons expos\'e  jusqu'{\`a}
pr{\'e}sent pr{\'e}serve la totalit{\'e} des  charges
supersym\'etriques  de la th{\'e}orie non compactifi\'ee
\cite{Vafa1,NSW}.  Cependant, les mod\`eles r\'esultants \`a quatre
dimensions,   par   exemple les supercordes  h\'et\'erotiques, ont
la supersym\'etrie d'espace-temps  et n'ont pas de fermions
chiraux.   La brisure partielle de la supersym\'etrie $ N=4$ vers
$ N=1$ est parmi les exigences  fondamentales pour construire des
mod\`eles de supercordes semi-r\'ealistes.
 Dans le but de remplir ces exigences,  il faudrait    compactifier les
supercordes  sur une vari\'et\'e complexe compacte   de
 Calabi-Yau  $\tilde X_6=CY_3$ avec  une courbure de Ricci nulle et ayant un
groupe d'holonomie  $SU(3)_H$ \cite{CHSW,DKV,Gren2}.  Ainsi
 cette compactification  permet de briser  le groupe de jauge de la
supercorde h\'et\'erotique  en identifiant  la connexion de spin du groupe
 $SU(3)_{H }$ de $X_3$ avec  la connexion de jauge  d'un sous groupe
$SU(3)_{YM }$ de l'un des deux facteurs  $E_8$  de la th\'eorie  $E_8 \t E_8$
\`a dix dimensions.  Le mod\`ele
r\'esultant \`a quatre dimensions  a  pour groupe de jauge $ E_6
\t E_8$.  \\ En g\'en\'eral,  les  vari{\'e}t{\'e}s de  Calabi-Yau
de dimension $n$ sont  des espaces   complexes, Kahl\'eriens,
compacts  ayant un tenseur de  Ricci nul et   un groupe
d'holonomie $SU(n)$. Il existe diff\'erentes fa\c cons  de
construire  ces vari{\'e}t{\'e}s. Nous citons:
\begin{itemize}
  \item { Orbifolds de $T^{2n}$.}
  \item  {Fibration elliptique sur une vari{\'e}t{\'e} complexe  $ B_{n-1}$ de
dimension $n-1$.}
  \item {Hypersurfaces dans les espaces projectives, o\`u plus
g{\'e}n{\'e}ralement dans les vari{\'e}t{\'e}s toriques.}
\end{itemize}
 \subsection{  Surface  K3}
 K3 est une vari{\'e}t{\'e} Kahl\'erienne compacte de dimension
  r\'eelle  $4$. Elle est  simplement connexe,  de courbure de Ricci nulle et
a
   un  groupe d'holonomie $ SU(2)$  assurant l'existence des spineurs
covariantiquement
   constants \cite{Besse,BPV,Borcea,Mat,Dona}.  Elle a la moiti{\'e} du nombre de
spineurs
    covariantiquement constants de $T^4$.  La compactification  sur  K3
pr{\'e}serve
      la moiti\'e des   charges supersym\'etriques  de  celles  de la
compactification sur
       $T^4$.  Cette  compactification  joue un r{\^o}le centrale dans les
conjectures
        de dualit{\'e}s des supercordes, dans la construction
g{\'e}om\'etrique des
         th{\'e}ories des champs supersym\'etriques d{\'e}velopp{\'e}e  par S.
Katz, P.
         Mayr et C. Vafa dans \cite{KMV}. Avant de consid\'erer
           la compactification des diff{\'e}rents mod\`eles de supercordes sur
K3,
            nous commen\c cons   tout d'abord par
            fixer  une r{\'e}alisation g{\'e}om{\'e}trique de  K3.  Nous
consid\'erons
              ici   la   r{\'e}alisation   de   $T^4\over Z_2$  avec
$T^4={R^4\o Z_4}$
               param\`etris\'e  par les coordonn\`ees r\'eelles  $x_i$  (o\`u
$i=1,\ldots,4$):
\be
x_i\equiv x_i+1. \ee
 En coordonn\'ees complexes  $z_1=x_1+ix_2$  et  $z_2=x_3+ix_4$, le tore $
T^4$  peut \^etre vu comme le plan complexe
$ \bf C^2$  quotient\'e par la sym\'etrie   $ Z_4$:
\be
z_k\equiv z_k+1,\quad z_k\equiv z_k+i,\quad k=1,2. \ee L'espace
$T^4\over Z_2$   est  obtenu \`a partir   de $T^4$ en imposant
la sym{\'e}trie
\be
 Z_2:  z_i \to -z_i,\quad i=1,2,
\ee
          renversant les  deux coordonn{\'e}es complexes   $z_i$ du tore
$T^4$. L'espace r{\'e}sultant
dit orbifold, pr{\'e}sente $16$ points singuliers correspondant  aux points
fixes de la sym\'etrie $ Z_2$.
Ce sont  les points  $( z_1^{(k)},z_2^{(k)}); k=1,\ldots,4$  avec  $
z_i^{(k)}=0,{1\o2}, {1\o2}i, {1\o2}+{1\o2}i$.\\
  Au voisinage de chaque  point singulier,   K3 se comporte comme l'espace
${\bf  C^2 }\o  Z_2$ de
 coordonn\'ees complexes $ (z_1,z_2)$ \cite{Sata,DHVW,Koba}. En effet, au
voisinage d'un point fixe $
(z_1,z_2)\equiv  (-z_1,-z_2)$,  nous pouvons choisir des coordonn\'ees
complexes $ u, v$ et $w$ qui sont invariantes sous la sym\'etrie $Z_2$
\be
 u=z_1^2,\quad  v=z_2^2,\quad w=z_1z_2.
\ee
 Cette nouvelle param\`etrisation d\'ecrit  un  mod\`ele local   singulier,
connu  sous le non de
 singularit{\'e} $A_1$ dans la classification des singularit{\'e}s des
surfaces complexes:
\be
 u  v=w^2.
\ee Cette \'equation  complexe peut aussi s'\'ecrire, par un choix
convenable des variables complexes $ u, v$ et $w$, sous la forme
\be
u^2 + v^2+w^2=0. \ee
 La partie r\'eelle de cette \'equation  n'est autre que  une  sph\`ere $S^2$
(2-cycle) de rayon z\'ero, donc d'aire nulle.
 Cette singularit\'e peut \^etre r\'esolue par une d\'eformation des
structures   de Kahler ou complexe   en changeant l'\'equation (3.2.6)  par
\be
u^2 + v^2+w^2=\chi. \ee
 G\'eom\'etriquement ceci correspond  \`a donner \`a la sph\`ere
pr\'ec\'edente une aire finie d\'etermin\'ee  par la
partie r\'eelle de $\chi$. Notons au passage que cette \'etude admet une
extension  remarquable en termes des
 singularit\'es de type ADE. Ils sont class\'es  comme  suit
\begin{equation}
\begin{array}{lcr}
A_n: \qquad x^2+y^2+z^{n+1}=0\\ D_n: \qquad y^2+x^2z+z^{n-1}=0\\
E_6: \qquad y^2+x^3+z^{4}=0 \\
E_7:\qquad y^2+x^3+xz^{4}=0 \\
E_8:\qquad  y^2+x^3+z^{5}=0.
\end{array}
\end{equation}
\\La homologie  du  $T^4\over Z_2$
contient les formes non twist\'ees  qui correspondent aux formes de l'espace
initial pr\'eservant  la sym\'etrie $Z_2$.
 Il y a 8 \'el\'ements r\'epartis comme suit
\begin{center}
\tabcolsep=16pt
\begin{tabular}{r|r}
   1&$h^{0,0}$=1\\
  $dz_1\wedge d z_2$&$h^{2,0}=1$\\
  $d{\bar z}_1\wedge d{\bar z}_2$&$h^{0,2}=1$\\
  $dz_i\wedge d{\bar z}_j$&$h^{1,1}=4$\\
  $dz_1\wedge dz_2\wedge d{\bar z}_1 \wedge d{\bar z}_2$&$h^{2,2}=1$\\
  \hline
  Total&8
\end{tabular}
\end{center}
o\`u $ h^{p,q}$ repr\'esente  le nombre des formes $(p,q)$ sur K3,
$p,q$ entiers  appartenant \`a la classe homologique $(p,q)$. En
plus de ces formes, nous avons  16   2-formes non triviales
correspondant aux  2-cycles non triviales qui   r\'esolvent   les
16  singularit\'es de  l'orbifold.  Ces formes  de type $H^{1,1}$
(K3) affectent uniquement  $h^{1,1}$ par un  facteur additif 16 de
sorte que $h^{1,1}=4+16=20$.   La cohomologie totale de K3 est
alors donn\'ee par le diamant de Hodge suivant
\cite{Vafa1,Todo,Morri,Aspin1}:\\
\def\m#1{\makebox[10pt]{$#1$}}
\begin{equation}
  {\arraycolsep=2pt
  \begin{array}{*{5}{c}}
    &&\m{h^{0,0}}&& \\ &\m{h^{1,0}}&&\m{h^{0,1}}& \\
    \m{h^{2,0}}&&\m{h^{1,1}}&&\m{h^{0,2}} \\
    &\m{h^{2,1}}&&\m{h^{1,2}}& \\ &&\m{h^{2,2}}&&
  \end{array}} \;=\;
  {\arraycolsep=2pt
  \begin{array}{*{5}{c}}
    &&\m1&& \\ &\m0&&\m0& \\ \m1&&\m{20}&&\m{1.} \\
    &\m0&&\m0& \\ &&\m1&&
  \end{array}}
\end{equation}
\subsection{  Espace des modules de  K3}
L'orbifold  K3  a $ h^{1,1}=20$  d{\'e}formations de Kahler
r\'eelles  et 20 d{\'e}formations complexes (40 param\`etres
r\'eels)  \cite{Todo,Morri,Seiberg,Aspin1,GP}. Ces derni\`eres sont  sp\'ecifi\'ees par le
choix de la 2-forme holomorphe  $\Omega_2$ de K3.   Tenant compte
du fait que pour chaque m\'etrique de  Ricci  plate de K3,  nous
pouvons  fixer un param{\`e}tre complexe, il reste   alors  $58 =
2 \times  19 + 20$  param{\`e}tres r{\'e}els. Le choix de la
m{\'e}trique  sur  K3 est donc param\`etris\'e par l'espace des
modules g\'eom{\'e}triques
\begin{equation}
 M^{geo}= {SO(19,3,{\bf R})\over {SO(19,{\bf R}) \times SO(3,{\bf R})}} \times
{\bf R}^+ ,
\end{equation}
o\`u  ${SO(19,3,{\bf R})\over {SO(19,{\bf R}) \times SO(3,{\bf
R})}}$ est   un r{\'e}seau pair autodual du type $ \Gamma^{19,3}$,
que l'on  a recontr\'e  lors de  la compactification toroidale des
supercordes h\'et\'erotiques  sur $T^3$ \cite{Vafa1}.  ${\bf R}^+$
correspond \`a un facteur d'{\'e}chelle global de la m{\'e}trique
de K3, qui peut  \^etre vu comme le volume de K3. Dans le cas
o{\`u} la vari\'et\'e  K3  est  r{\'e}alis{\'e}e comme une
fibration elliptique de la sph{\`e}re $ S^2$, qui peut \^etre vue
localement comme
\be
K3=T^2\times S^2, \ee
 l'espace des modules est de dimension complexe 18 plus  2   param{\`e}tres
r\'eels
  correspondant  aux classes de Kahler de la base et  de la fibre. Cet espace
est isomorphe \`a
\be
M^{geo}= {SO(18,2,{\bf R})\over {SO(18,{\bf R}) \times SO(2,R)}}
\times {\bf R}^+\t {\bf R}^+. \ee
\\
 Les modules  g\'eom\'etriques  de K3 ne suffisent  cependant pas  {\`a}
d\'ecrire la
 compactification de la th{\'e}orie des supercordes sur cet espace.
  Il  faut  ajouter  les   valeurs moyennes du tenseur antisym{\'e}trique
   $ B_{\mu\nu} $ du secteur  ${\bf NS-NS}$,  celles des  tenseurs
antisym{\'e}triques de
     ${\bf R-R}$ dans le cas  des mod\`eles  type II, et celles des   champs de
      jauge dans le cas de la th{\'e}orie de type I et les supercordes
h\'et\'erotiques,
       ainsi  le dilaton qui correspond {\`a} la constante de couplage de la
supercorde
          $g_s= e^{\phi}$.
\subsection{Vari\'et\'es de Calabi-Yau de dimension 3 } Ce sont des
espaces complexes  ayant   un groupe d'holonomie $SU(3)$
pr{\'e}servant  le quart  $({1\o4})$  du nombre initial des
charges supersym\'etriques \`a dix dimensions \cite{GSW,Pol1,Vafa1}. Ces
vari{\'e}t{\'e}s sont compactes de dimension trois complexes qui
restent  les candidats les plus  probables pour connecter les
mod\`eles de supercordes {\`a} notre {\'e}chelle d'observation \`a
quatre  4 dimensions. Comme pour K3, ces vari\'et\'es admettent
aussi une construction  comme  orbifold  de $T^6$. L'id\'ee de
cette construction est de consid\'erer $T^6$ comme le produit $
T^2\t T^2\t T^2$ et  quotienter ensuite  par une transformation
$Z_3$ simultan\'ee sur chaque $T^2$ comme suit: \bea
 x_1+ix_2\to  w_1(x_1+ix_2),\nn\\
 x_3+ix_4\to  w_2(x_3+ix_4),\nn\\
x_5+ix_6\to  w_3(x_5+ix_6), \eea o\`u $w_i,i=1,2,3$ sont des
\'el\'ements de $Z_3$ satisfaisant $ w_i^3=1$.  L'espace
r\'esultant est  une vari\'et\'e  de Calabi-Yau avec $ 3^3=27$
points  fixes.  Ces points  singuliers  doivent \^etre r\'esolus
afin de rendre la vari\'et\'e de Calabi-Yau  r\'eguli\`ere. Notons
que  dans cette r\'ealisation, le nombre  de Hodge $h^{1,1}$ \'egaul  \`a 36
 ($h^{1,1}=36$) r\'eparti  comme suit:  une contribution $h^{1,1}_{nt}=9$
prevenant du secteur non twist\'e et correspond au choix de la
forme  $dz_i\wedge d{\bar z}_j$. L'autre contribution
$h^{1,1}_{t}=27$ provient du secteur twist\'e. Ce nombre 27
correspond  aux  essoufflements   des   point fixes par  des
sph\`eres  $S^2$ de contibution $h^{1,1}=1\t 27=27$.\\
 Notons qu'il existe  deux autres  fa\c cons pour r\'ealiser  une vari\'et\'e
de Calabi-Yau  de dimension  trois  $CY_3$.\\
 (1) Elle   peut \^etre d\'efinie comme une  hypersurface dans l'espace
projectif  $ \bf  P^4$  de coordonn\'ees  homog\`enes $ (z_1,z_2,z_3,z_4,z_5)$.
 La condition de Calabi-Yau
exige que l'\'equation alg\'ebrique de $CY_3$ est  donn\'ee par un
polyn\^ome homog\`ene  $ W_5(z_i)$  de degr\'e cinq
\be
 W_5(z_i)=a_1 z_1^5+ a_2 z_2^5+a_3 z_3^5+a_4 z_4^5+a_5
z_5^5+a_6z_1z_2z_3z_4z_5+\ldots
\ee Les d\'eformations complexes sont donn\'ees par les
coefficients complexes  $a_i$ dont le nombre est donn\'e par les
126 param\`etres    moins les 25 param\`etres de la sym\'etrie
$U(5)$, qui est en accord avec  $  h^{2,1}=126-25=101$. Quant aux
d\'eformations de Kahler, nous avons une  seule  ($ h^{1,1}=1$)
qui correspond  au volume  de $X_3$.  (2) $CY_3$  peut \^etre vue
comme la fibration elliptique     sur une base complexe  de
dimension deux  $B_2$ param\`etris\'ee  par  deux coordonn\'ees
complexes $ (z_1,z_2)$. Localement nous avons
\be
X_3=T^2\t B_2. \ee
    L'\'equation alg\'ebrique de ce mod\`ele  est donn\'ee par
\be
y^2= x^3+ f (z_1,z_2)x+g(z_1,z_2), \ee o\`u  $(x,y)$ sont  deux
variables complexes param\`etrisant   la courbe elliptique $T^2$.
Notons que l'espace des param\`etres de ce mod\`ele   d\'epend de
la base $B_2$. Ce type de vari\'et\'es sont \'etudi\'ees dans le cadre de la
compactification de la th\'eorie
-F \cite{Vafaf}.
\section{ Compactification des  Supercordes  sur K3}
\subsection{ Supercorde  IIA sur K3} Dans la
compactification du supercorde de type IIA sur K3,   nous obtenons
une th{\'e}orie supersym\'etrique $N = 2$ {\`a} 6 dimensions.
Puisque $ b_1(K3) = b_3(K3) = 0$, les champs de Ramond ne
g\'en\`erent pas  des  degr{\'e}s de libert{\'e}
suppl{\'e}mentaires sur K3.  Alors que      les valeurs moyennes
du tenseur antisym\'etrique $ B_{\mu\nu} $ peuvent {\^e}tre
mesur{\'e}es par  $ b_2(K3)=22$, o\`u
 $b_k= \sum \limits _{p+q=k} h^{p,q}$.
Le spectre complet des {\'e}tats non massifs de cette th{\'e}orie
est donn\'e par  l'espace des modules \cite{Vafa1}
\begin{equation}
 M^{IIA}= {SO(19,3,{\bf R})\over {SO(19,{\bf R}) \times SO(3,{\bf
R})}}\times{\bf R}^+\times{\bf R}^{22} \times R^+,
\end{equation}
 qui peut \^etre \'ecrit sous la forme
\begin{equation}
 M^{IIA}= {SO(20,4,{\bf R})\over {SO(4,{\bf R}) \times SO(20,{\bf R})}}
\times SO(1,1)
\end{equation}
 o\`u $SO(1,1)$  correspond  au dilaton. Cet espace des modules
 est identique {\`a} celui de la supercorde h{\'e}t{\'e}rotique
compactifi{\'e}e sur un tore $T^4$ \cite{NSW}.
Cette coincidence n'est pas forfuite  c'est  une  premi\`ere  indication   de
l'exsitence   d'une  dualit\'e entre la
  supercorde  IIA et celles  h\'et\'erotiques. Nous reviendrons sur cette
connection entre ces mod\`eles de supercordes
 dans le dernier paragraphe de ce chapitre.  La sym\'etrie $ O(20,4,{\bf Z})$
contient la sym\'etrie g\'eom\'etrique $ O(19,3, {\bf Z})$
 de K3. Elle d\'ecrit des dualit\'es perturbatives des th\'eories de
supercordes analogues \`a la dualit\'e-T  des compactifications toroidales.

\subsection{ Supercorde  IIB sur K3}

Dans le cas du  mod\`ele IIB, le secteur {\bf  NS-NS } donne  les
m{\^e}mes  modules que  la th\'eorie IIA,  par contre le secteur
{\bf R-R} donne des modules suppl{\'e}mentaires
\cite{Vafa1}. Le nombre total de ces modules est
r{\'e}sum{\'e} dans le tableau suivant:
\begin{center}
\tabcolsep=16pt
\begin{tabular}{r|r}
   M\'etrique&58\\
 Champ $B$&22\\
  Dilaton&1\\
Axion &1\\
  2-forme&22\\
  4-forme&1\\
  \hline
  Total&105
\end{tabular}
\end{center}
Ces champs scalaires se combinent  pour donner  l'espace total
des modules de  IIB sur K3
\begin{equation}
 M^{IIB}= {SO(21,5,{\bf R})\over {SO(21,{\bf R}) \times SO(5,{\bf R})}}.
\end{equation}
 Cet espace admet une identification sous  les transformations  du groupe $
SO(21,5,{\bf Z})$.\\
 Notons que les compactifications de type II sur K3 permettent de r{\'e}duire
la dimension de l'espace-temps {\`a} 6
dimensions tout en conservant la moiti\'e des  charges supersym\'etriques.
Les mod\`eles de ces th{\'e}ories {\`a}
 D = 4 peuvent {\^e}tre obtenus par une compactification suppl{\'e}mentaire
sur un espace de dimension deux. Par exemple  dans le cas
 de  la compactification sur un tore $T^2$, on obtient une th\'eorie
supersym{\'e}trique  $N=4$  \`a   quatre dimensions \cite{{100},{101}}.
 \subsection{ Mod\`eles  de supercordes $N = 1$ sur K3}
Rappelons que les secteurs bosoniques de ces mod\`eles  $N = 1$
sont $ g _{\mu \nu}$, $ B _{\mu \nu}$,  $\phi$ et  $ A^\mu=A^\mu
_a T^a$,  o\`u $\{T^a\}$ sont les g{\'e}n{\'e}rateurs du groupe de
jauge $ SO(32)$ ou $ E_8\times E_8$. Pour les champs $ g _{\mu
\nu}$,  $ B _{\mu \nu}$ et $\phi$,      nous avons la m{\^e}me
situation que pour la compactification  du secteur {\bf NS-NS} des
mod\`eles  de type II sur K3.  Concernant le champ de jauge
$A^\mu$, nous pouvons choisir une configuration non nulle sans
briser la supersym\'etrie. Le point est que si l'on consid{\`e}re
des configurations des champs de jauge sur la surface  K3 satisfaisant la
condition d'auto-dualit{\'e}:
 \begin{equation}
F = \ast F,
\end{equation}
alors elles pr{\'e}servent le m{\^e}me nombre de charges
supersym\'etriques que  la m{\'e}trique sur K3.  Ces
configurations, qui   correspondent aux   instantons standards de
la th{\'e}orie de jauge, doivent {\^e}tre prises en
consid{\'e}ration car  dans  th{\'e}ories $ N = 1$, nous devons   satisfaire
l'{\'e}quation suivante
 \begin{equation}
  dH= Tr F\wedge F -Tr R\wedge R,
\end{equation}
o{\`u}  $H = dB$ est le champ fort du champ antisym{\'e}trique  $
B_{\mu\nu}$.  Notons qu'en l'absence de singularit{\'e} pour le
champ $H$, nous avons  la contrainte suivante
\begin{equation}
   \int _ {K3}dH=0.
\end{equation}
Dans ce cas, le nombre des instantons  de jauge, qui est
reli{\'e} {\`a} la caract{\'e}ristique d'Euler $ \chi (K3)$, est
24. Pour le groupe de jauge  $E_8 \times  E_8$, ce nombre peut
{\^e}tre r\'eparti entre  les  deux facteurs $E_8$ comme $(12+n ,
12-n)$.  Notons qu'il   est possible de consid{\'e}rer aussi des
configurations de vide telles que
\begin{equation}
   \int _ {K3}dH=m.
\end{equation}
Dans ce cas, nous avons $24-m$ instantons. L'espace des modules
complet de la compactification des th{\'e}ories $N = 1$ contient
les modules d\'eterminent    la g{\'e}om{\'e}trie de la surface  K3, les
valeurs moyennes du  champ antisym{\'e}trique $ B_{\mu\nu}$,   le
dilaton ainsi les modules d{\'e}crivant le choix du fibr\'e de
jauge,   qui peut \^etre vu comme l'espace des modules des
instantons sur la surface K3 \cite{Vafa1}.
 \section{ Compactifications sur des Calabi-Yau de dimension 3}
\subsection {Compactification des mod\`eles    $N = 2$} La
compactification des mod\`eles  de supercordes de type II  sur
n'importe quelles  vari{\'e}t{\'e}s de Calabi-Yau    {\`a} trois
dimensions complexes  conduit {\`a} une  th{\'e}orie
supersym\'etrique $N = 2$  {\`a} 4 dimensions \cite{sen10,Witten3,CL,CGH,Sen2,NSV}.
Dans cette th\'eorie, nous avons trois   types de supermultiplets
contenant   les champs  scalaires non massifs:
\begin{itemize}
  \item { Un  multiplet vectoriel  $N=2$  contenant   deux scalaires
r{\'e}els, deux fermions de Weyl et un boson vecteur  $( 0^2, {{1\over2}^2},
1)$.}
  \item {Un  hypermultiplet $N=2$ contenant    quatre  scalaires r{\'e}els et
deux fermions de Weyl  $( 0^4, {{1\over2}^2})$.}
  \item { Un supermultiplet gravitationnel  $( g_{\mu\nu}, A_{\mu},
\psi_{A\mu},\psi_A).$ }
\end{itemize}
 En g\'en\'eral, l'espace des modules des th\'eories  supersym\'etriques  $
N=2$  se   factorise  en un produit
\be
 {\cal M}_V\otimes  {\cal M}_H,
\ee o\`u   ${\cal M}_V$ est  une vari\'et\'e Kahl\'erienne  de
dimension $2N_V$ param\`etris\'ee  par les scalaires  des  $N_V$
multiplets vectoriels  et $ {\cal M}_H$ est une vari\'et\'e
hyper-Kahl\'erienne de dimension $4N_H$,  correspondante   aux
$N_H$   hypermultiplets.\\ Dans le cas du mod\`ele  IIA sur une
vari{\'e}t{\'e} de Calabi-Yau   $CY_3$, nous avons:
\begin{itemize}
  \item  {
$h_{1,1}(CY_3)$   param\`etres complexes sp\'ecifiant   la
structure de  Kahler  de $CY_3$  ainsi    le choix  du champ
antisym\'etrique $ B_{\mu\nu}$ du secteur {\bf  NS-NS}.} \item {
$h_{2,1}(CY_3)$ modules complexes  d\'eterminant  la d\'eformation
de la structure complexe de $ CY_3$.} \item { $h_{2,1}(CY_3)$
param\`etres complexes correspondant aux choix de la 3-forme $
C_{\mu\nu\rho}$.} \item { $h_{0,3}+h_{3,0}=2$      param{\`e}tres
r{\'e}els  provenant  du champ  3-forme $ C_{\mu\nu\rho}$, plus 2
modules r{\'e}els    correspondant au    dilaton et un champ
scalaire   dual du champ antisym\'etrique $ B_{\mu\nu}$ \`a
quatre dimensions.} \end{itemize}
  Ces quatre champs scalaires   forment un
hypermultiplet suppl\'ementaire.  Par cons\'equent,   nous avons:
\begin{equation}
\begin{array}{lcr}
h_{1,1}(CY_3) \qquad  { multiplets  \;vectoriels}\\ h_{2,1}(CY_3)+1
\qquad   hypermultiplets.
\end{array}
\end{equation}
Si nous consid{\'e}rons la  supercorde    IIB   sur  une
vari\'et\'e $CY^*_3$, au lieu de la supercorde IIA  sur $CY_3$, nous
avons $ 2 h_{1,1}(CY^*_3)$   param{\`e}tres complexes  sp\'ecifiant
la d{\'e}formation de Kahler de $CY^*_3$ et le champ $ B^{{\bf
NS}}_{\mu\nu}  $ et   $ \tilde B^{ {\bf R}}_{\mu\nu} $
respectivement, 2 modules r{\'e}els provenant du dilaton et
l'axion de la th{\'e}orie IIB. Nous avons aussi $h_{2,1}(CY^*_3)$
param{\`e}tres complexes  param\`etrisant  la structure complexe
de $CY^*_3$.   Alors   que  la 4-forme $ D_{\mu\nu\rho \lambda}$
donne $h_{2,1}(CY^*_3)$  bosons vectoriels.  Ces modules
peut {\^e}tre arrang\'es    comme suit:
\begin{equation}
\begin{array}{lcr}
h_{2,1}{(CY^*_3)} \qquad  {multiplets \; vectoriels}\\ h_{1,1}{(CY^*_3)}
+ 1 \qquad hypermultiplets.
\end{array}
\end{equation}
 Nous remarquons   que les r{\^o}les des structures complexes et  de  Kahler
des  vari{\'e}t{\'e}s de Calabi-Yau $CY_3$ et $CY^*_3$  sont interchang\'ees :
\begin{equation}
\begin{array}{lcr}
h_{2,1}(CY_3)  \to h_{1,1}(CY^*_3) \\ h_{1,1}(CY_3) \to h_{2,1}(CY^*_3).
\end{array}
\end{equation}
  Cette  sym{\'e}trie est connue sous  sym{\'e}trie miroir.   Sous cette
transformation, le mod\`ele  type
 IIA sur $CY_3$ est {\'e}quivalente au mod\`ele type IIB  sur   la   varit{\'e}t{\'e} miroir $CY^*_3$ \cite{Vafa1}.  Notons au passage que
 cette sym{\'e}trie  est utlis\`ee
 dans la construction g{\'e}om{\'e}trique
 des th\'eories  supersym\'etriques  \`a  quatre dimensions, notamment dans la
d\'etermination des solutions exactes
 de la branche de Coulomb du mod\`ele type IIA\cite{KMV,BS}
\subsection{Compactification des mod\`eles $N = 1$} La
compactification des supercordes $N = 1$ (h\'et\'erotiques $E_8\t
E_8$, $SO(32)$ et type I) sur des  Calabi-Yau de dimension trois
conduit {\`a} differents  mod\`eles  $N = 1$ {\`a} quatre
dimensions.  Comme  pour la compactification des supercordes $N =
1$ sur la surface  K3, nous devons choisir une configuration des champs de
jauge de sorte que
\begin{equation}
  Tr F\wedge F =-{1\over2}Tr R\wedge R.
\end{equation}
Il se trouve qu' il y a plusieurs fa{\c c}ons de satisfaire cette
contrainte. Une fa\c con particuli{\`e}rement int\'eressante   est
de faire coincider  un sous  groupe  $SU(3)_{YM}$ du groupe de
jauge $G $ avec le groupe d'holonomie  $SU(3)_H$ de la
vari{\'e}t{\'e} de Calabi-Yau tridimensionnelle. Pour  un des
deux  facteurs  $E_8$ de la th{\'e}orie de supercorde $E_8 \times
E_8$, nous avons la brisure suivante:
\begin{equation}
  E_8  \longrightarrow E_6 \times SU(3).
\end{equation}
 La repr\'esentation  adjointe 248, du groupe $ E_8$  se d\'ecompose en termes
des repr\'esentations  $(m,n)$ de $ E_6\t SU(3)_K$  de  la  mani\`ere
suivante:
\be
248=(78,1)+(1,8)+(27,{\overline{ 3}})+({\overline{ 27}},3).
 \ee
 La mati\`ere  charg\'ee est dans les  repr\'esentations  $27$ et ${\overline{
27}} $.
  Le nombre 27 est donn\'e par $h^{(1,1)}$ et celui de ${\overline{ 27}}$ est
   donn\'e par $h^{(1,2)}$.  Notons qu'il existe aussi plusieurs champs
neutres qui
     correspondent  \`a la d\'eformation  de
      la structure de Kahler (complexe) de la vari\'et\'e de Calabi-Yau
tridimensionnelle.
      En plus de ces champs  scalaires neutres,  nous avons d'autres
param\`etrisant
      l'espace des modules de la connexion    de jauge sur la vari\'et\'e de
Calabi-Yau
      tridimensionnelle.
\chapter{  Solitons  en   th{\'e}ories des supercordes}
\pagestyle{myheadings}\markboth{\underline{\centerline{\textit{\small{Solitons
 en   th{\'e}ories des
supercordes}}}}}{\underline{\centerline{\textit{\small{Solitons  en
th{\'e}ories des supercordes}}}}}
  Afin de mieux illustrer  l'aspect non perturbatif  de la th\'eorie des
supercordes,  nous commen\c cons tout d'abord
par rappeler certains r\'esultats  concernant   la th\'eorie des champs hors
du r\'egime perturbatif.   A  quatre  dimensions,
les ph\'enom\`enes non perturbatifs  peuvent  \^etre obtenus  par   la
recherche des solutions  non triviales des \'equations de
 mouvement des champs classiques.  Ces m\'ethodes non perturbatives ont
connus un d\'eveloppement remarquable apr\`es la d\'ecouverte
 des instantons de t'Hooft et  de Belavin et al \cite{P,T}.  Ces instantons
concernent  les  solutions  auto-duales et antiauto-duales des
 \'equations de Yang-Mills dans un espace Euclidien \`a quatre dimensions.
 \begin{equation}
F = \pm  \ast F,
\end{equation}
o\`u $F$ est  la courbure de la connexion de jauge.\\ Les
instantons contribuent aux diff\'erentes quantit\'es physiques par
des termes en $ e^{- { 1 \o g^2}}$, $g$ etant la constante de
couplage,  et correspondent \`a des transitions par effet tunnel
entre les vides classiques de la th\'eorie.\\
 Un deuxi\`eme type  de configuration des champs  jouant  un r\^ole important
en physique non perturbative correspond  aux
 solutions statiques  \`a \'energie finie des \'equations de mouvement. Ces
solutions, dites  solitons, sont au contraire
 ind\'ependantes  du temps et   localis\'ees dans l'espace. Leurs masses
varient  comme $ 1 \o g^2$, et  \'echappent donc au
spectre perturbatif $( g\to 0)$.  En  th\'eorie des champs non perturbative,
ces solutions correspondent  \`a des nouvelles particules
qui ne sont pas cr\'ees par les champs fondamentaux de la th\'eorie. \\
 Ayant discut\'e les solitons  en th\'eorie  des champs
   nous retournons maintenant \`a \'etudier l'analogue de ces objets  en
th\'eorie des
    supercordes.  Pour commencer  notons que l'{\'e}tude du spectre des
     {\'e}tats solitoniques dans les th{\'e}ories de supergravit\'e {\`a} 10
dimensions
       r\`ev\`ele l'existence  des d'objets {\'e}tendus \`a   $p$ dimensions
spatiales
         appel{\'e}s  $p$-branes \cite{PP}. Ces objects charg{\'e}s sous des
tenseurs
         antisym{\'e}triques $ (p+1)$-formes apparaissent  de fa\c con
naturelle en
           th\'eorie des supercordes  et jouent   un  r\^ole
consid{\'e}rable dans
           la compr{\'e}hension des dualit{\'e}s des mod\`eles de
supercordes \cite{Vafa1,PP,W1,BSV1,BSV2,DKP,SW3}.

\section{Branes  et  tenseurs antisym{\'e}triques}

Avant de donner une description explicite des $ p$-branes, il est
int{\'e}ressant de rappeler que le spectre des {\'e}tats de masse
nulle des th{\'e}ories de supercordes se compose de:   le dilaton
$ \phi$, dont la valeur moyenne d{\'e}finit la constante de
couplage de la supercorde  $ g_s=e^{-\phi}$,  le graviton $g_{\mu
\nu}$ de spin 2 et un certain nombre de tenseurs
antisym{\'e}triques de jauge d{\'e}pendant du mod\`ele  de
supercorde en question.  Ces tenseurs antisym\'etriques
g{\'e}n{\'e}ralisent la notion du potentiel vecteur $ A_\mu$ {\`a}
un tenseur antisym{\'e}trique  {\`a} $ p+1$  indice $A_{p+1}$,
(o\`u $(p+1)$-forme),
\be
A_{p+1}=A_{{\mu_1}\ldots{\mu_{p+1}}}dx^{{\mu_1}}\ldots
dx^{{\mu_{p+1}}}. \ee
 Ce champ est invariant sous la transformation de jauge suivante
  \be
A_{p+1}  \longrightarrow  A_{p+1}+d \lambda_p, \ee o\`u $d$ est la
derivation ext\'erieure  $(d^2=0)$  et $\lambda_p$ est  une $
p$-forme.  Le champ fort invariant de jauge est donn\'e par
\be
F_{p+2}=dA_{p+1}, \ee satisfaisant   l'\'equation de Maxwell
\be
d^*F_{p+2}=0. \ee Les   objets charg{\'e}s sous ces tenseurs
antisym\'etriques de jauge $ A_{p+1}$ sont appel\'es  $p$-branes.
Pour  mieux illustrer  la chose, nous  rappelons  que  la
dynamique d'une particule ponctuelle coupl{\'e}e {\`a} une
1-forme $ A_ \mu$ implique le couplage:
\begin{equation}
  S_A =\int d\tau A_\mu {d X^\mu \over d\tau}.
\end{equation}
La charge {\'e}lectrique associ\'ee   {\`a} un champ de jauge
$A_\mu $   dans un espace de dimension $d$ est mesur{\'e}e par le
flux du champ {\'e}lectrique {\`a} travers  une sph{\`e}re
$S^{d-2}$
\begin{equation}
  Q_E =\int \limits _ { S^{d-2}} \ast F_d,
\end{equation}
o\`u $ F_d = dA$. La charge magn{\'e}tique est  determin\'ee  de
mani\`ere identique  \`a  l'aide de   la dualit{\'e} de
Poincar{\`e} $dA \longrightarrow  \ast dA $  par
\begin{equation}
  Q_M =\int \limits _ { S^{d}}  F.
\end{equation}
De mani\`ere  analogue  on associe \`a chaque champ  de jauge $
A_{{\mu_1}\ldots{\mu_{p+1}}}$, \` a $ p+1$ indices,   un objet
\'etendu  \` a   $p$ dimensions spaciales et   dont le chemin au
cours du temps occupe  un  volume de dimension $p+1$ dans
l'espace-temps.  Cet objet appel\'e  $p$-brane  g\'en\'eralise la
notion de particule ponctuelle $(p=0) $ et la corde $ (p=1)$  \`a
des objets de dimension interne d'ordre sup\'erieure $p>1$
\cite{Vafa1}.
  Par cons\'equant  une $p$-brane se couple \`a   $
A_{{\mu_1}\ldots{\mu_{p+1}}}$  par un couplage
\be
\int A_{{\mu_1}\ldots{\mu_{p+1}}}dx^1\ldots dx^{p+1} \ee
 g\'en\'eralisant  les lignes de Wilson des th\'eories de jauge et nous avons
les  r\'esultats  suivants r\'ealis\'es  dans  un espace-temps \`a $d$
dimensions:\\
 $(\imath$)  La charge {\'e}lectrique  associ\'ee  au champ $A$ \` a $p+1$
indices est mesur{\'e}e par:
\begin{equation}
  Q_E =\int \limits _ { S^{d-(p+2)}} \ast dA.
\end{equation}
$(\imath \imath$) La charge magn{\'e}tique est d{\'e}finie \`a
l'aide de  la dualit{\'e} de Poincar\'e  par:
\begin{equation}
  Q_M =\int \limits _ { S^{p+2}}  dA.
\end{equation}
 $(\imath \imath \imath$)  La dualit{\'e} {\'e}lectrique -magn{\'e}tique en
dimension $d$  change donc  une $p$-brane  en une   $q$-brane avec
\begin{equation}
p+q=d-4
\end{equation}
En particulier   pour une  3-forme  coupl\'ee  \`a  2-brane nous
avons
\begin{equation}
 S_{A_3} =\int d^3\sigma  \xi ^{\alpha\beta \gamma}\partial _{\alpha}X^\mu
\partial _{\beta}X^\nu \partial _{\gamma }X^\rho.
\end{equation}
 L'\'etude des supercordes \`a dix dimensions  montre  qu'il existe deux
 types des branes:  {\bf NS-NS} -branes  charg{\'e}es sous les tenseurs de
 jauge  du secteur  {\bf NS-NS}, et D$p$ -branes  charg{\'e}es sous les
 champs antisym{\'e}triques du secteur  {\bf R-R}.
Nous avons la classification suivante de type de D-branes que l'on
recontre en th\'eorie des supercordes
\begin{center}
\begin{tabular}{|c|c|c|c|c|c|}
 \hline
   & Type IIB & Type IIA & H\'et\'erotique  & H\'et\'erotique  & Type I \\
    &  &  & $E_8\times E_8$ & $SO(32)$ &  \\ \hline
  Type de corde & ferm\'ee & ferm\'ee & ferm\'ee & ferm\'ee & ouverte  \\
   & &  &  &  &  (et ferm\'ee) \\ \hline
  Supersym\'etrie& $N=2$ & $N=2$& $N=1$ & $N=1$ & $N=1$ \\
   de l'espace-temps & chirale  &  non chiral   &  &  &  \\ \hline
  Sym\'etrie de jauge &- & - &$E_8\times E_8$  & $SO(32)$ & $SO(32)$ \\ \hline
  D-branes & -1,1,3,5,7 & 0,2,4,6 &- & - & 1,5,9\\ \hline
\end{tabular}
\end{center}

\section{ La physique  D-branes }

 Les D$p$-branes  peuvent \^etre vue comme des hypersurfaces de dimension $p$
imerg\'ees dans l'espace-temps,
 ils  apparaissent comme des objets infiniment massifs \`a faible couplage et
leurs masses
 $m={1 \o g_s^2}$,  o\`u $g_s$ est la constante de couplage de corde.  En
r\'egime  du couplage fort   $(g_s\to \infty)$,
ces objects deviennent  non massifs et dominent la dynamique non perturbative
de la th\'eorie.  De point de vue  th{\'e}orie des
 supercordes,  les D $p$-branes sont des  hypersurfaces  sur lesquelles des
supercordes ouvertes  sont attach{\'e}es \cite{Pol1,Pol2}.
La notion de D $p$-brane, traduit le fait  que les
extr{\'e}mit{\'e}s des  supercordes ouvertes  satisfont   les
conditions de  bords de {\bf  Dirichlet}:\\
\begin{equation}
 \partial _\tau  X^\mu (\sigma=0,\pi)=0, \quad \mu =p+1,\ldots, 9.
\end{equation}
 suivant les directions  transveres;  ainsi elles  r\'epondent    aux
conditions de {\bf  Neumann} sur les  $(p+1)$  directions  longitidinales,
c.\`a.d
 \begin{equation}
 \partial _\sigma X^\mu (\sigma=0,\pi)=0,\quad  \mu =0,\ldots, p.
\end{equation}
Contrairement \`a  la th\'eorie des supercordes  ouvertes libres
o\`u les modes se propagent dans l'espace-temps \`a dix
dimensions, la sym\'etrie de  Lorentz  $SO(1,9)$ \`a dix
dimensions   se brise en  $SO(1,p) \times SO(9-p)$ sur  le volume
d'univers  de la $p$-brane. Les fluctuations des champs des
supercordes ouvertes  d\'ecrivent  alors  la  dynamique  de  la
D$p$-brane.  De plus les modes de masse nulle  des supercordes
ouvertes attach\'ees {\`a} la D-brane  correspondent  aux
degr{\'e}s de libert{\'e} de la D$p$-brane.  Puisque   le spectre
de masse nulle des supercordes ouvertes correspond  au potentiel
vecteur $ A^{\mu}$, la condition d'attachement  sur   les D$
p$-branes implique  que ce  champ se propage  uniquement  sur le
volume d'univers de dimension ($p+1$):
\be
 A^{\mu}= \alpha_{-1}^\mu |0,k> ,\quad  \mu=0,\ldots, p,
\ee en pres\`ence des   $9-p$ champs scalaires $A_i$, du point de
vue du volume d'univers de la D$p$-brane,
\be
A^{i}= \alpha_{-1}^i |0,k> , \quad i= p+1, \ldots , 9. \ee
\subsection{Action des $p$-branes}
En pres\`ence du champ de jauge $A^\mu $, l'action d\'ecrivant la
dynamique de la D$p$-brane est donn{\'e}e par l'action de Born
-Infeld \cite{Pol1}
\be
{\cal S_{BI}}=\int d^{p+1} e^{-{\phi\o 2}} {\sqrt {det
(g_{\mu\nu}+2\pi\a'F_{\mu\nu}}}), \ee
 o{\`u} $ F=dA$ est la courbure du champ de jauge $A$ et  $g$ repr{\'e}sente
la  m{\'e}trique induite  sur
le volume d'univers.  La tension de la Dp-brane est donn{\'e}e   par
\be
T_p={e^{-{\phi}}\o (\a')^{{p+1\o2}}}. \ee En pr\'esence d'un champ
antisym{\'e}trique  $B_{\mu\nu}$  du secteur de {\bf NS-NS},
l'action de Born -Infeld peut {\^e}tre g\'en\'eralis{\'e}e  de la
mani\`ere suivante
\be
{\cal S_{BI}}=-T_p\int d^{p+1} e^{-{\phi\o 2}}{ \sqrt {det
(g_{\mu\nu}+2\pi\a'{\cal F}_{\mu\nu})}}-iT_p\int A_{p+1}, \ee
 o\`u  ${\cal F_{\mu\nu}}$ est la  courbure g{\'e}n{\'e}ralis{\'e}e donn\'ee
par
\be
{\cal F_{\mu\nu}}=2\pi\a'F_{\mu\nu}- B_{\mu\nu}=2\pi
\alpha'(\p_\mu A_\nu-\p_\nu A_\mu)-B_{\mu\nu}. \ee Cette action
peut  \^etre  complet\'ee  par  l'adjonction des  fermions dans le
but   d'avoir   une th{\'e}orie  supersym{\'e}trique  sur le
volume d'univers.  Si nous  consid{\'e}rons que la D $p$-brane est
approximativement plate,  la dynamique de ses fluctuations  est
d\'ecrite par  une  th{\'e}orie de  Maxwell ordinaire $U(1)$ \`a
$(p+1)$ dimensions avec $ 9-p$ champs scalaires.  Apr{\`e}s  cette
addition des champs fermioniques  $ \psi$, l'action {\`a}  faible
{\'e}nergie  peut  \^etre  \'evalu\'ee explicitement  par la
r{\'e}duction dimensionnelle de la th{\'e}orie de Maxwell $U(1)$
supersym\'etrique $N = 1$ \`a dix dimensions
\begin{equation}
S= {1\over g^2_{YM}}\int d^{10}x (-{1\over 4} F_{\mu\nu}F^{\mu\nu}
+{1\over 2} \bar\psi \Gamma^\mu\partial _\mu \psi )
\end{equation}
 vers une th\'eorie  {\`a} $(p+1)$ dimensions. Cependant   cette th\'eorie,
invariante sous
la moiti\'e des charges  supersym\'etriques, d\'ecrit
   la dynamique d'une seule  D$p$-brane. \\
Une  g\'en\'eralisation de cette  analyse  peut {\^e}tre \'etendue
au cas  de  plusieurs D$p$-branes parall\`eles. Ceci  conduit  \`a
la  th{\'e}orie de Yang-Mills supersym{\'e}trique. Pour une
configuration  des  supercordes ouvertes  attach{\'e}es sur deux
D$ p$-branes diff{\'e}rentes, la masse des \'etats fondamentaux de
ces supercordes  ouvertes  est donn\'ee par
\be
M={L \o \alpha'},
 \ee qui  est proportionnelle  \`a leur
{\'e}longation $L$. Ces \'etats deviennent non massifs  lorsque
les  deux D $p$-branes  coincident $(L=0)$. Dans  le cas de $N$
D$p$-branes confondus,  nous avons $ N^2$  champs de jauge des
supercordes ouvertes $ A_{ij}^\mu $.  Ceci conduit \`a  une
th{\'e}orie de Yang-Mills non ab{\'e}lienne $ U(N)$ sur le volume
d'univers de dimension $p+1$ dont  l'action  est d\'efinit par
\be
{\cal  S}_p^N=T_p \;str\; \int \limits _{W_{p+1}}d^{p+1}
e^{-{\phi\o 2}}( F_{\mu\nu}^2+2F_{\mu I}^2+F_{IJ}^2)
\ee
 o{\`u}
$
F_{\mu\nu}=\p_\mu A_\nu -\p_\nu A_\mu +\[ A_\nu,
A_\mu\]$,   $
 F_{\mu I}=\p_\mu X_I  +\[ A_\mu, X_I\]$ et $
F_{IJ}=\[ X _I, X_J]$ avec $A_{\mu}$,  $X_I$ sont des matrices de
$U(N)$ \footnote{ Nous signalons qu'il y a d'autres extensions
associ\'ees avec les groupes  SO  et SP}.
  Notons que dans le cas o\`u le potentiel est nul
\be
Tr\; {\[X^I,X^J\]}^2=0, \ee les matrices de position commutent et
peuvent \^etre simultan\'ement diagonalis\'ee.  Dans ce cas nous
avons la version abeli\`enne  et nous retrouvons la notion de
position individuelle de chaque brane.
\section{  D-branes  et les espaces de Calabi-Yau }
Partant de  $p$-branes  nous pouvons  r\'eduire  certaines de ses
dimensions  internes
 en  les enroulant autour  des directions compactes. L'enroulement d'une
D$p$-brane  sur des cycles compacts  ($C_r$) de dimension
 $ r$  signifie que  son volume d'univers contient $r$ directions
compactifi\'ees   conduisant \`a une  D$(p-r)$-brane dans
l'espace-temps non compact. En fait,  on peut  d{\'e}composer  le potentiel de
jauge $A_{p+1}$ en produit de  deux potentiels  l'un
 dans les directions compactes de dimension $ r$ et l'autre   $ (p+1-r)$
-forme dans les directions non compactes; soit alors
\be
A^{p+1}\to w_r\wedge A^{p+1-r}, \ee
 o\`u $w_r$ est  une $r$-forme harmonique sur la vari\'et\'e compacte
\cite{Vafa1}. \par
Dans  la compactification sur des vari{\'e}t{\'e}s de Calabi-Yau,
l'enroulement d'une   D$p$-brane sur un cycle arbitraire  brise
toutes les supersym{\'e}tries, ces cycles sont  dites
supersym{\'e}triques. On dinstigue   deux cat{\'e}gories des
cycles  supersym\'etriques  dans  une  vari\'et\'e de Calabi-Yau
$CY_d$ de  dimension complexe $d$:\\ ($\imath$) Les
cycles de type A, de dimension r{\'e}elle $ d$, sont tels que la
forme holomorphe $(d,0)$ est proportionnelle  {\`a} la  forme
volume induite par la m\'etrique  de la vari\'et\'e  $CY_d$  sur
les cycles compactes. La compactification  sur ces cycles  conduit
\`a  des D $p$-branes de type A.\\ ($\imath\imath $) Les cycles de
type B  sont  les sous vari\'et\'es complexes holomophes de la
vari\'et\'e $CY_d$. Ils  conduisent  \`a D $p$-branes de type B.\\
Dans le contexte de la compactification des th{\'e}ories de
supercordes II   vers les   dimensions inf\'erieures,  nous
distinguons deux cas particuliers int\'eressants:\\ $\spadesuit$
D$p$-brane  enroul{\'e}e sur un  $p$-cycle
 $C_p$  de dimension r\'eelle $p$  et de volume $V_p$ donne  une  particule
($p-p=0$-brane) charg{\'e}e sous un champ de
 jauge 1-forme ($ A_\mu$), obtenu par la d\'ecomposition de $(p+1)$-forme en
terme d'une
  $p$-forme harmonique $w_p$ sur $C_p$,  dans l'espace-temps non compact
\be
\int_{C_p}A^{p+1}\to\int d\tau A.
\ee La masse $M$ de  D0-brane
r\'esultante  est proportionnelle au volume $ V_p$
\be
M\sim V_p.
\ee
Dans la limite o\`u $V_p$  tend vers z\'ero   nous
obtenons un nouveau  \'etat  non perturbatif de masse nulle. Cette
id\`ee est  \`a la base  de la  construction g\'eom\'etrique des
th\'eories de superYang-Mills \`a partir  de la th\'eorie des
supercordes  II \cite{KKV}.\\ $\spadesuit$  D$p$-brane avec un
volume d'univers enclidien  de dimension $(p+1)$ enroul\'ee sur un
$(p+1)$-cycle conduit  \`a des instantons de la th\'eorie des
supercordes. Ces objets g\'en\'erent des corrections non
perturbatives par une action $ e^{S} $, o\`u $S=tension
(brane)\;\;\t \; volume\; d'univers$.
\chapter{Dualit\'es  en Th\'eorie   des Supercordes}
\pagestyle{myheadings}
\markboth{\underline{\centerline{\textit{\small{Dualit\'es  en
Th\'eorie   des
Supercordes}}}}}{\underline{\centerline{\textit{\small{Dualit\'es
en Th\'eorie   des Supercordes}}}}}
\section{ Dualit{\'e}  en  th{\'e}orie des supercordes}
La  sym\'etrie de dualit{\'e} a conduit {\`a} une r{\'e}volution
dans  la conception des th{\'e}ories de supercordes  puisqu'elle a
permis de voir les cinq mod\`eles de supercordes comme des
manifestations  d'une seule th\'eorie \cite{Witten1}, dite th\'eorie-M.  En
g{\'e}n{\'e}ral, une dualit{\'e} entre deux th{\'e}ories est une
transformation reliant  leurs  {\'e}tats \cite{Vafa1}. La dualit{\'e}
a \'egalement  la propri\'et\'e  de transformer   un probl{\'e}me
difficile {\`a} r{\'e}soudre dans une th{\'e}orie en probl{\`e}me
facile {\`a} r{\'e}soudre dans la th{\'e}orie duale, en
particulier elle permet de transposer l'\'etude de    la limite
du couplage fort  d'une th{\'e}orie de supercordes \`a celle  du
couplage faible dans la th{\'e}orie duale, comme c'est le cas des
th\'eories h\'et\'erotiques  et type I.
\\
 Dans ce chapitre nous nous int\'eressons   {\`a} la dualit{\'e}
  entre la  supercorde  h\'et\'erotique et la supercorde  type  IIA \`a six
et   \`a  quatre dimensions.
  Ces dualit\'es ont \`a la base de plusieurs r\'esultats
  obtenus durant les quelques derni\`eres  ann\'ees.
\section{ Sym{\'e}trie de dualit{\'e}:  D\'efinitions}
Nous avons  cinq  mod\`eles  de supercordes  critiques  {\`a} 10
dimensions: IIA,  IIB,  type I $SO(32)$, h\'et\'erotique $SO(32) $
et h\'et\'erotique  $E_8 \times  E_8$. La compactification  de ces
mod\`eles donne plusieurs  mod\`eles de supercordes dans les
dimensions inf{\'e}rieures. Chacune de ces mod\`eles poss\`ede   un espaces
des modules
param\'etris\'e  par les modules suivants:
\begin{itemize}
  \item {  La constante de
couplage de la supercorde  $g_s=e^{<\phi>}$, o\`u  $<\phi>$ est  la
valeur moyenne du dilaton  dans le vide.}
  \item { Les modules
g{\'e}om{\'e}triques de la vari\'et\'e    compacte $X$
dont le nombre provient des diff\'erents  choix  possibles de la
m\'etrique.  Pour un espace de Calabi-Yau de dimensions $n$, Ce nombre est
donn\'e  par les  d{\'e}formations de
Kahler $h^{1,1}(CY_n)$ et les d{\'e}formations
complexes  $h^{n-1,1}(CY_n)$ de $CY_n$.}
  \item { Les  valeurs  moyennes des champs
antisym{\'e}triques  des secteurs  {\bf NS-NS}, {\bf R-R} et
des champs de jauge. }
\end{itemize}
 Ces  trois  types  de valeurs moyennes
param\`etrisent  l'espace des modules de la th{\'e}orie
compactifi\'ee \cite {Vafa1} sur une vari\'et\'e $X$.  Dans  la  r{\'e}gion   de
l'espace  des
modules o\`u   la  constante de couplage est faible, la
th{\'e}orie perturbative est relevente. Alors que   dans la
r{\'e}gion o\`u la constante de couplage est  forte,  c'est
plut\^ot   le  r\'egime   non perturbatif  qui est dominant.  Un
exemple   de sym{\'e}trie de dualit{\'e} est  celle qui transforme
une r{\'e}gion  perturbative d'une th{\'e}orie \`a  une r\'egion
non perturbative d'autre  th\'eorie et vice versa. Quand la
sym{\'e}trie de dualit{\'e} relie  les deux r{\'e}gions de la
m\^eme th{\'e}orie,   nous disons  que   la th{\'e}orie  est
auto-duale (par example,  mod\`ele IIB).
Pour faciliter la lecture, nous avons  jug\'e utile  de revoir
bri\`evement les types de sym\'etries des dualit\'es que nous
rencontrons en th\'eorie des supercordes.
\subsection {Dualit{\'e}-T}
Comme nous avons vu  dans   la compactification des mod\`eles de   supercordes
sur un cercle de rayon $R$.  La  dualit{\'e}-T
consiste {\`a} faire la transformation  $ R \longrightarrow
{\alpha' \over R}$.  Cette sym{\'e}trie  se  g{\'e}n{\'e}ralise
vers le groupe  $SO(d,d,\bf Z)$  lors de  la compactification
toroidale  sur le tore  $T^d$.    Cette dualit{\'e} relie   la
r\'egion de faible couplage de deux th\'eories diff\'erentes.  Par
exemple, le mod\`ele  IIA sur un cercle de rayon $R$ est dual   au mod\`ele
IIB  sur un cercle de rayon ${1 \o R}$.   De la  m\^eme fa\c con
les deux  mod\`eles de supercordes h\'et\'erotiques sont
\'equivalentes \`a neuf dimensions. A cause  de  cette
sym{\'e}trie de dualit{\'e},  les cinq mod\`eles  de supercordes \`a dix
dimensions
se r{\'e}duisent \`a trois mod\`eles  distincts  \`a neuf dimensions:  type
II,  type I et supercorde h\'et\'erotique.   Il
est naturel de chercher d'autres relations entre  ces trois mod\`eles
de supercordes. Sachant que  la dualit{\'e}-T  intervient dans une
limite perturbative. Nous nous attendons \`a retrouver  un lien
entre ces trois mod\`eles de mani{\`e}re  \`a \'etendre la
dualit\'e perturbative vers une dualit\'e  non perturbative. Pour
cela  nous allons \'etudier autre type de dualit{\'e}.
\subsection {  Dualit\'e-S}
 C'est une transformation de dualit{\'e} reliant  deux  r{\'e}gimes
de couplages diff{\'e}rents. Elle est  donn\'ee
  par  l'inversion du couplage $g_s$,
\be
g_s \longrightarrow {1 \over g_s}. \ee
 Cette dualit\'e  nous permet de  d\'ecrire  le couplage fort (faible) d'une
th\'eorie  \`a
 l'aide du  r\'egime \`a
  couplage faible (fort) de sa th\'eorie duale.   La dualit\'e-S va  au del\`a
  de la
 dualit{\'e} {\'e}lectrique-magn{\'e}tique apparaissant
 en    th{\'e}orie de  Yang-Mills \`a  quatre dimensions.  Notons qu'il exite
un   autre
 type de  dualit{\'e} dite { \it  dualit{\'e}-U}.
Cette derni\`ere est une combinaison de  la dualit\'e-T et la dualit\'e-S.\\

\section{Dualit\'e faible-fort couplage \`a dix dimensions}
 Dans ce paragraphe,  nous d\'ecrivons    deux exemples {\`a} dix dimensions:
Le premier
correspond \`a  la dualit{\'e}-S
$( Sl(2,Z))$   de la th{\'e}orie de type IIB, tandis que le deuxi{\`e}me
exemple est
 celui de  la dualit{\'e} h\'et\'erotique SO(32)-type I SO(32).
\subsection{ Auto-dualit\'e du  mod\`ele IIB }
 Rappelons que le supermultiplet gravitationnel du mod\`ele type IIB
  \`a 10 dimensions contient le graviton $ g_{\mu\nu}$, deux gravitino
  $ \psi_{\mu,\a}^{\dot{\a}}$  deux tenseurs antisym{\'e}triques
  $B_{\mu\nu}$, $ \tilde B_{\mu\nu}$, un champ scalaire  complexe  $\tau
=\chi +ie^{-\phi}$, dont la
partie imaginaire est identifi\'ee  avec la constante de couplage  de la
th{\'e}orie des supercordes
et  une  quatre forme $ D_{\mu\nu\rho\lambda}$ auto-duale $ dD=*dD$.
 La dualit\'e-S est une transformation  particuli\`ere du  groupe $ Sl(2,{\bf
Z})$
 agissant sur le param\`etre $\tau=\chi+ie^{-\phi}$ comme \cite{Witten1,Schwarz,PW}
\be
\tau \to {a\tau+b\over c\tau+d}; \quad  a,b,c,d \in {\bf Z},\quad
ad-bc=1, \ee
 o\`u $\chi$ correspond {\`a} la valeur moyenne  de l'axion   de {\bf R-R } du
mod\`ele type  IIB.
 D'autre part  la    paire  $ (B_{\mu\nu},  \tilde B_{\mu\nu} )$ se transforme
sous la sym\'etrie
$Sl(2,{\bf Z})$ comme un doublet
\be
{B_{\mu\nu}\choose \tilde B_{\mu\nu}}\to \left(\matrix{ a&b\cr
c&d\cr}\right){B_{\mu\nu}\choose \tilde B_{\mu\nu}}, \ee
 alors que  la m\'etrique et la   quatre forme restent invariantes. Dans le
cas $\chi=0$,
 nous voyons que $g\to {1\o g}$ et  $B \to \tilde B$ et  $\tilde B \to - B$
c'est  la sym\'etrie de
 dualit{\'e}-S.
Cette sym\'etrie   \'echange la supercorde fondamentale charg\'ee
sous $  B_{\mu\nu}$ en D1-brane (D-corde) charg\'ee  sous  $
\tilde B_{\mu\nu}$.  Une fa\c con  de voir cette \'equivalence est
au niveau  de  leurs  actions  sur la surface d'univers.   En
effet,  la partie bosonique de la supercorde fondamentale du
mod\`ele  type  IIB
\be
S_F={1\over \a'}\int d\sigma^2(\p_aX^i)^2, \ee
 peut \^etre  identifi\'ee avec l'action de la D-corde. Cette action est
d\'ecrite dans la jauge du
c\^one de lumi\`ere par  un champ de jauge $A_\alpha$ et  huit champs
scalaires $X^i$ \`a deux dimensions.  En particulier, elle est
obtenue par  la r\'eduction dimensionnelle  de la th\'eorie  de super
Yang-Mills \`a  dix dimensions vers   deux dimensions
\be
S_D={1\over g_s\a'}\int d\sigma^2\{F^2_{\a\beta}+(\p_aX^i)^2\},
\ee o\`u $F^2_{\a\beta}$ est le champ  fort de  $A_\a$. Ce dernier  ne
contenant    pas des degr\'es de libert\'e de jauge   et  qui
conduit par suite \`a  l'\'equivalence cis mentionn\'e.   Notons
au passage  que la  {\bf NS}5-brane est \'egalement
identifi\'ee \`a  la D5-brane du secteur  de {\bf R-R} du mod\`ele   type IIB
alors la D3-brane est invariante.
\subsection{ Dualit{\'e}  type I- h{\'e}t{\'e}rotique  SO(32) }
A dix dimensions, la th{\'e}orie  super Yang Mills avec la sym\'etrie de jauge
$SO(32)$  a deux
r{\'e}alisations  diff\'erentes  l'un provenant du mod\`ele
de type I  et  l'autre de  la supercorde   h{\'e}t{\'e}rotique
$SO(32)$. Quoique, ces  deux  mod\`eles  ont le m{\^eme} spectre
des \'etats de masse nulle $ ( g_{\mu \nu}, B_{\mu \nu}, \phi,
A_{\mu})$, elles ne peuvent {\^e}tre reli\'ees trivialement \`a
cause de leurs  propri\'et\'es perturbatives  compl\`etement
diff\'erentes.  Ce qui nous am\`ene  \`a conjecturer l'exsitence
d'une  dualit\'e faible-fort couplage: $$ g_{Het }
\longrightarrow {1 \over g_{I}}.$$
 Cette transformation  identifie le tenseur  $ B_{\mu\nu}$  de la supercorde
h\'et\'erotique
 $SO(32)$ avec le tenseur $ \tilde B_{\mu\nu}$ du mod\`ele  type I,  $SO(32)$.
 Donc la supercorde h\'et\'erotique, charg\'ee sous $B_{\mu\nu}$,  peut \^etre
identifi\'ee avec la D1-brane
du mod\`ele type I,
  charg\'ee sous $\tilde B_{\mu\nu}$. Les champs de jauge associ\'es  \`a la
sym\'etrie $SO(32)$ de la supercorde
h\'et\'erotique correspondent  pr\'ecis\'ement aux champs de la surface
d'univers de la D1-brane. En particulier,
  ces champs correspondent
aux modes de masses nulles des supercordes ouvertes dont  les extr\'emit\'es
sont attach\'ees  \`a la  D1-brane. Finalement
 la  {\bf NS} 5-brane h\'et\'erotique   peut \^etre   \'egalement
identifi\'ee \`a la  D5-brane du  mod\`ele type I \cite{PW}.

Dans  cette  section, nous avons  d\'ecrit
deux exemples de la  dualit{\'e} faible-fort couplage. Connaissant
le r\'egime \`a faible couplage,  nous pouvons alors
d{\'e}terminer   la limite de couplage fort  des  mod\`eles  IIB,
type I et  h{\'e}t{\'e}rotique $SO(32)$. Une  question  naturelle
qui se pose \`a ce niveau: Quelle est  la dynamique  \`a fort
couplage des th{\'e}ories  IIA et la supercorde
h{\'e}t{\'e}rotique $E_8\times E_8$?   Il se trouve que  la
dynamique \`a fort couplage de ces  deux derni\`eres  a une
origine diff\'erente et son interpr\'etation  n\'ecessite  d'aller
au del\`a  de la dimension dix  d'espace-temps, vers onze dimensions \cite{Witten1}.
\section{Dualit{\'e} type II- h\'et\'erotique}
 \subsection{  Type IIA sur K3/ h\'et\'erotique sur $T^4$}
 Apr\`es  compactification sur un cercle,  la  dualit\'e ne laisse que trois
mod\`eles de supercordes
 apparemment distinctes: type II (IIA  est \'equivalente IIB),  mod\`ele type
I   $SO(32)$ et la
 suprecorde  h\'et\'erotique ($SO(32)$ ou $ E_8\t E_8$ ).   De plus les  deux
modeles  $SO(32)$  sont aussi identifi\'ees par
  la dualit\'e faible-fort couplage. La question  qui se pose a ce niveau et
de  voir s'il
 existe  \'egalement une description duale dans laquelle les th\'eories $N=1$
et $N=2$  soient reli\'ees?.  La  r\'eponse
  est positive; en  effet,   le cas
le plus \'etudi\'e  est  donn\'e  par  la dualit\'e  de  la
th\'eorie de type IIA sur la surface  complexe  K3   et  la th{\'e}orie h\'et\'erotique
sur $T^4$ \cite{HT,DKV,Vafa1,Kir}.  Ces  deux th\'eories  pr{\'e}sentent
le m{\^e}me espace des modules:
 \begin{equation}
SO(1,1) \times { SO(20,4,{\bf  R}) \over SO(20,{\bf  R})
SO(4,{\bf  R})},
\end{equation}
o\`u $SO(1,1)$  correspond \`a  la constante de couplage de la
corde,  et le quotient ${ SO(20,4, {\bf  R}) \over SO(20,{\bf  R})
SO(4,{\bf  R})}$ correspond  \`a l'espace  des modules du r\'eseau
pair auto-dual de Narain    $\Gamma _{20,4}$,   d\'efinissant la
compactification de la supercorde h\'et\'erotique sur $T^4$,  ou
encore aux modules de la compactification de la supercorde  type IIA  sur K3. Notons
que  la th{\'e}orie des supercordes h\'et\'erotiques poss{\'e}dent
une sym{\'e}trie de jauge non ab{\'e}lienne,  alors que la
th{\'eorie de type IIA sur K3 ne  poss{\`e}de que  des champs de
jauge ab{\'e}liens   provenant du secteur R-R. Ces
champs  correspondent  soit au  champ de jauge  1- forme
(D0-branes),  soit  \`a   la r{\'e}duction  du  tenseur
antisym{\'e}rique 3- forme (D2-branes) sur les deux cycles de K3.
Quoique,  ces deux th\'eories ont des groupes de jauge
diff\'erents, la conjecture de dualit{\'e}  IIA sur K3 et
h\'et\'erotique sur $T^4$  a pass\'ee avec succ\`es plusieurs
tests en particulier en utilisant les  D-branes. La sym{\'e}trie
non ab\'elienne  de la supercorde h\'et\'erotique  est associ\'ee
en  IIA \`a la sym\'etrie  des  points singuliers  de l'espace des
modules de K3.   Lorsqu'un ou plusieurs 2-cycles s'annulent, les
D2-branes de la th\'eorie de type IIA  enroul{\'e}es autour d'eux
g{\'e}n{\'e}rent des particules  de jauge de masse  nulle portant
une sym{\'e}trie de jauge non ab{\'e}lienne d{\'e}termin\'ee par
la matrice d'intersection des 2-cycles de K3 \cite{Aspin2,Aspin3,Ber,KMP}.
De plus,  les singularit{\'e}s de K3 ont \'et\'e  class\'ees par
les singularit{\'e}s de type ADE.  De cette propri\'et\'e, il en
d\'ecoule entre autre que l'on peut identifier  les \'etats de la
supercorde h\'et\'erotique tels que ceux du spectre
$(P^2_L,P^2_R)=(0,2)$   du r\'eseau de Narain  avec le r\'eseau
d'homologie de   la vari\'et\'e K3 singuli\`ere
 \footnote{ On associe \`a chaque racine
simple des alg\`ebres de Lie ADE  un 2-cycle de K3.  Ainsi   les
matrices d'intersections des 2-cycles de K3 aux celles de Cartan
des alg\'ebres de Lie.}.
\subsection{Dualit\'e  h\'et\'erotique -type IIA \`a  quatre  dimensions}
 La dualit\'e $N=2$ h\'et\'erotique -type IIA \`a six dimensions que nous
venons de d\'ecrire  peut  \^etre  \'etendue   \`a   quatre dimensions tout en
conservant  la totalit\'e ou  la moiti\'e  des supersym\'etries \`a six
dimensions \cite{HT2,AFIQ,Kachru,KLM,Klemm,KLT,OV,SV,Sen4}. Cette  \'etude est bas\'ee sur   l'argument  adiabatique de Vafa
et Witten \cite{VW}. Ce dernier  consiste \`a faire une compactification des
deux th\'eories   sur un m\^eme espace de dimension r\'eelle deux. \\
{ \bf {Argument adiabatique}}.\\ En  g\'en\'eral  partant  d'une
th\'eorie $A$  compactifi\'ee sur une vari\'et\'e $K^A$ duale \`a
une th\'eorie $B$  compactifi\'ee sur une vari\'et\'e $K^B$,
\begin{center}
A sur $K^A$ $\sim $ B sur  $K^B$,
\end {center}
  et compactifions ces deux   th\'eories  sur un   m\^eme espace $ \tilde M$,
nous produisons une nouvelle paire des th\'eories duales dans des dimensions
inf\'erieures:
\begin{center}
A sur ($K^A \t \tilde M )$ $\sim $ B sur  $(K^B \t \tilde M).$
\end {center}
Ce programme   qui est connu sous  l'argument adiabatique  stipule
que   la dualit\'e entre les th\'eories  ci-dessus peut \^etre vu
comme \'etant une dualit\'e entre les fibres $K$ en  tout point de
l'espace
 $\tilde M$. Consid\'erant
  la dualit\'e h\'et\'erotique - type IIA \`a six dimensions et
  utilisant
  l'argument adiabatique  on peut d\'eduire  d'autres  dualit\'es \`a  4
dimensions en prenant  $ \tilde M_2=T^2$ ou $S^2$  avec  $N=4 $ et
$N=2$ respectivement.


\newpage

\begin{thebibliography}{99}
\pagestyle{myheadings} \pagestyle{myheadings}
\markboth{\underline{\centerline{\textit{\small{
BIBLIOGRAPHIE}}}}}{\underline{\centerline{\textit{\small{BIBLIOGRAPHIE}}}}}
\bibitem{GSW}
 M. Green, J. Schwarz and E. Witten, {\em  Superstring Theory},
vol 1 and 2, Combridge University Press, 1987.
\bibitem{Pol1}
  J. Polchinski, {\em  String theory}, vol and 2, Cambridge  University Press, 1999.
\bibitem{Vafa1}
C. Vafa, {\em Lectures on Strings and Dualities}, {\tt hep-th/9702201}.
\bibitem{Rand1}  S. Randjbar-Daemi, {\em  Introduction to Chiral  Anomalies},  lectures
presented at Introductory School on String Theory,  ICTP, Trieste, Italy,
(1998).
\bibitem{Na1}   K. S. Narain,  {\em Toroidal compactification and heterotic string},
lectures presented  at  Introductory School on String Theory,  ICTP, Trieste,
Italy (1998).
\bibitem{GH} C. Gomez and  R. Hernandez, {\em Fields, strings and branes},
{\tt hep-th/9711102}.\\ C. Gomez, lectures presented at  the Workshop on
Noncommutative Geometry, Superstrings and Particle Physics. Rabat
-Morocco, (May 11-12 2001).
\bibitem{Pol2}J. Polchinski, {\em What is String Theory?}, {\tt  hep-th/9411028}.
\bibitem{Kir}  E. Kiritsis, {\em Introdution to Superstring Theory}, {\tt hep-th/9709062}.\\
E. Kiritsis, {\em Introduction to Non-perturbative string theory}, {\tt
hep-th/9708130}.

\bibitem{Poly1} A. M. Polyakov, Phys. Lett.{ \bf B103}(1981)207-211.
\bibitem{Poly2} A. M. Polyakov, Phys. Scr. {\bf 15} (1987)191.
\bibitem{Ram} P. Ramond, {\em Dual Theory for fermions}, Phys. Rev.
D3(1971)2415. \\
            A. Neveu, J. H. Schwarz, Nucl. Phys. {\bf B31} 86.
\bibitem{Ven} G. Venieziano, {\em An introduction to dual models of
strong interactions and their physical motivations}, Phys. Rep. {\bf  C9}
199.\\ J. H. Schwarz, {\em  Dual resonance theory}, Phys. Rep.
           {\bf C8}(1973)269.\\   A. Neveu, J. Schwarz, Nucl. Phys. {\bf B31}(1971)86.\\
  P. Ramond, Phys. Rev. {\bf D3} (1971)2415.
\bibitem{GSO1}
 F. Gliozzi, J. Scherk and D. Olive, Phys. Lett
{\bf  B65}(1976)282.
\bibitem{GSO2}
    F. Gliozzi, J. Scherk and D. Olive, Nucl. Phys.
             {\bf B122} (1977)282.
\bibitem{BPZ}
A.A. Belavin, A. M. Polyakov and A. B. Zamolodchikov, J. Stat.
Phys. {\bf 34}, (1984), 763; Nucl. Phys. {\bf B241}(1984)333.
\bibitem{FQS}
 D. Friedan, Z. Qiu and S. Shenker, Phys. Rev. Lett.{\bf 52}(1984)1575.
\bibitem{FV} S. Fubini and G. Veneziano,  Ann.
Phys. {\bf 63}(1970)12.
\bibitem{NST} A. Neveu,  J. H. Schwarz and C. B. Thorn, Nucl. Phys.
            {\bf  B35}(1971)529.
\bibitem{ON} M. O'Lougnlin, K.S. Narain, {\em  Non-Perturbative Aspects Of
Supersymmetric String Theories}, lectures  presented at  Introductory School on
String Theory,   ICTP, Trieste, Italy (1998).
\bibitem{Gren1} B. M. Grenne, {\em  Aspects of  D- Geometry},  lectures  presented at
Spring school on superstring theories and related
 matters, ICTP, Trieste, Italy, (1999).
\bibitem{Di} P. D.  Vecchia,  {\em Large  N gauge theories and ADS/CFT
Correspondence},  lectures  presented at Spring school on superstring theories
and related
 matters, ICTP, Trieste, Italy, (1999).
\bibitem{Anto} I. Antoniadis, { \em Mass Scales in String and M-Theory},  lectures
presented at Spring school on superstring theories and related
 matters, ICTP, Trieste, Italy, (1999).
\bibitem{Malda1}  J. Maldacena, {\em ADS/CFT Correspondence}, lectures  presented at
Spring school on superstring theories and related
 matters, ICTP, Trieste, Italy, (1999).
\bibitem{Kach}
S. Kachru,  {\em Warped  Brane Worlds and  Hierarchy Problems}, lectures
presented at  Spring School on superstrings and related matters ,
ICTP, Trieste, Italy, (2000).
\bibitem{Malda2}  J. Maldacena, {\em  The large N limit of Field Theories and Gravity},
lectures  presented at Spring school on superstring theories and related
 matters, ICTP, Trieste, Italy, (2000).
\bibitem{Malda3}  J. Maldacena, {\em The Gravity /Field theory Correspondance},
lectures  presented at Spring school on superstring theories and related
 matters, ICTP, Trieste, Italy, (2000).
\bibitem{Kuta} D. Kutasov, {\em  Introduction to little string theory}, lectures
presented at Spring school on superstring theories and related matters, ICTP,
Trieste, Italy, (2001).
\bibitem{CHSW}   P. Candelas,  G. Horowitz, A. Strominger, E. Witten, {\em Vacuum
configurations for superstrings}, \newblock  Nucl.  Phys  {\bf B258} (1985) 46.
\bibitem{NSW} K. Narain, M. H.  Sarmadi, E. Witten, {\em   Note on the Toroidal compatification of Heterotic string
Theory},
\newblock  Nucl.  Phys  {\bf B279} (1987) 369-379.
\bibitem{Yau}  S-. Yau
\newblock {\em Calabi Yau Conjecture and Somme New results in Algebraic
 Geometry}
\newblock  Pro. Natl.Acad.Sci {\bf B74} (1977) 1798-1799.
\bibitem{Gren2} B. R. Grenne, {\em  String Theory on Calabi Yau Manifolds,
 {\bf TASI} lectures 1996},  { \tt hep-th/9702155}.
\bibitem{Thei} S. Theisen,  {\em Introduction to Calabi-Yau manifolds}, Lectures
presented at Spring school on superstring theories and related matters, ICTP,
Trieste, Italy, (2001).
\bibitem{Dou} M. Douglas, lectures presented at Spring school on
 superstring theories and related matters, ICTP, Trieste, Italy, (2001).\\
\bibitem{Witten1}
E.~Witten,
 {\em  String  Theory Dynamics in Various Dimensions}, Nucl. Phys. {\bf
443}(1995)184, {\tt  hep-th/9507121}.
\bibitem{Schwarz}
J.~H. Schwarz,
\newblock {\em An SL(2,Z) Multiplet of Type IIB Superstrings},
\newblock Phys. Lett. {\bf B360} (1995) 13-18, {\tt  hep-th/9508143}.
\bibitem{HT}
C.~Hull and P.~Townsend,
\newblock {\em Unity of Superstring Dualities},
\newblock Nucl. Phys. {\bf B438} (1995) 109--137, {\tt hep-th/9410167}.
\bibitem{DKV}
L.~J. Dixon, V.~Kaplunovsky, and C.~Vafa,
\newblock {\em On Four-Dimensional Gauge Theories from Type-II Superstrings},
\newblock Nucl. Phys. {\bf B294} (1987) 43-82.
\bibitem{KLMVN}
  A. Klemm, W. Lerche, P. Mayr , C.  Vafa, N.  Warner,  Nucl. Phys {\bf
B477}(1996)746.
\bibitem{Vafaf}
C. Vafa, {\em  Evidence for F-Theory},  Nucl.Phys. {\bf B469}
(1996)403-418,  {\tt hep-th/9602022}.

\bibitem{KKV}
S. Katz,A. Klemm and C. Vafa,  Nucl. Phys {\bf B497}(1997)
173-195.
\bibitem{KMV} S. Katz, P. Mayr and C. Vafa, {\em  Mirror symmetry and exact solution
of 4d  N=2 gauge theories I},  Adv,  Theor.  Math. Phys {\bf 1}(1998)53.
\bibitem{SW1} N. Seiberg and Witten, {\em  Electric-Magentic duality, Monople
Condensation and confinement in N=2 supersymetric Yang-Mills Theory},  Nucl.
Phys. {\bf B 426} (1994) 19.
\bibitem{SW2} N. Seiberg and  E.Witten, {\em  Monople, Duality  and Chiral Symmetry
Breaking in N=2 Supersymmetric QCD},    Nucl. Phys. {\bf B 431} (1994) 484.
\bibitem{HW} A. Hanany and E. Witten,  Nucl. Phys  {\bf B492}  (1997)152-190,
{\tt hep-th/9611230}.
\bibitem{Witten2}  E. Witten,  Nucl. Phys  {\bf B500}(1997)3-42, {\tt hep-th/9703166}.
\bibitem{belhaj1}
A.  Belhaj,  {\em On Geometric Engineering of Supersymmetric Gauge
Theories}, the proceedings of the Workshop on Noncommutative
Geometry, Superstrings and Particle Physics. Rabat -Morocco,
(16-17 June 2000), {\tt hep-ph/0006248}.
\bibitem{BS}
A. Belhaj, E.H Saidi, {\em
  Toric Geometry, Enhanced non Simply laced Gauge Symmetries in Superstrings
and F-theory Compactifications},  {\tt hep-th/0012131}.
\bibitem{Sentac}
N. Berkovits, A. Sen, B. Zwiebach, {\em Tachyon Condensation in Superstring Field Theory},  Nucl.Phys. {\bf B587} (2000) 147-178, {\tt hep-th/0002211}.

\bibitem{Sentac1}A.  Sen, B. Zwiebach,
{\em  Tachyon condensation in string field theory}, JHEP {\bf 0003} (2000) 002,  {\tt hep-th/9912249}.

\bibitem{GHMR} D. Gross, J. Harvey, E. Martenic and R. Rohm, Nucl. Phys.
           {\bf   B256}(1985)253.
\bibitem{Gren3} B. Grenne, {\em String theory on  Calabi-Yau manifolds},
{\tt hep-th/9702155}.
\bibitem{Besse}
A.~L. Besse,
\newblock {\em Einstein Manifolds},
\newblock Springer-Verlag, Berlin, 1987.
\bibitem{BPV}
W.~Barth, C.~Peters, and A.~Van~de Ven,
\newblock {\em Compact Complex Surfaces},
\newblock Springer, 1984.
\bibitem{Borcea}
C.~Borcea,
\newblock {\em Diffeomorphisms of a K3 surface},
\newblock Math. Ann. {\bf 275} (1986) 1-4.
\bibitem{Mat}
T.~Matumoto,
\newblock {\em On Diffeomorphisms of a K3 Surface},
\newblock in M.~{Nagata et al}, editor, ``Algebraic and Topological Theories
  --- to the memory of Dr. Takehiko Miyaka'', pages 616--621, Kinukuniya,
  Tokyo, 1985.
\bibitem{Dona}
S.~K. Donaldson,
\newblock {\em Polynomial Invariants for Smooth Four-Manifolds},
Topology {\bf 29} (1990) 257-315.
\bibitem{Sata}
I.~Satake,
\newblock {\em On a Generalization of the Notion of Manifold},
\newblock Proc. Nat. Acad. Sci. USA {\bf 42} (1956) 359-363.
\bibitem{DHVW}
L.~Dixon, J.~A. Harvey, C.~Vafa, and E.~Witten,
\newblock {\em Strings on Orbifolds},
\newblock Nucl. Phys. {\bf B261} (1985) 678-686,
\newblock and {\bf B274} (1986) 285--314.
\bibitem{Koba}
R.~Kobayashi,
\newblock {\em Einstein-K{\"a}hler V-Metrics on Open Satake V-Surfaces with
  Isolated Quotient Singularities},
\newblock Math. Ann. {\bf 272} (1985) 385-398.
\bibitem{Todo}
A.~Todorov,
\newblock {\em Applications of K{\"a}hler--Einstein--Calabi--Yau Metric to
  Moduli of K3 Surfaces},
\newblock Inv. Math. {\bf 61} (1980) 251-265.
\bibitem{Morri}
D.~R. Morrison,
\newblock {\em Some Remarks on the Moduli of {K3} Surfaces},
\newblock in K.~Ueno, editor, { \em lassification of Algebraic and Analytic
  Manifolds}, Progress in Math.~{\bf 39}, pages 303-332, Birkh{\"a}user,
  1983.
\bibitem{Aspin1} P. Aspinwall, {\em  K3 surfaces and String Duality}, {\tt  hep-th/961117}.
\bibitem{Seiberg}
N.~Seiberg,
\newblock {\em Observations on the Moduli Space of Superconformal Field
  Theories},
\newblock Nucl. Phys. {\bf B303} (1988) 286-304.
\bibitem{GP}
B.~R. Greene and M.~R. Plesser,
\newblock {\em Duality in CY Moduli Space},
\newblock Nucl. Phys. {\bf B338} (1990) 15-37.

\bibitem{sen10}
A.~Sen,
\newblock {\em Strong-Weak Coupling Duality in Four-Dimensional String Theory},
\newblock Int. J. Mod. Phys. {\bf A9} (1994) 3707-3750, {\tt hep-th/9402002}.

\bibitem{Witten3}
E.~Witten,
\newblock {\em New Issues in Manifolds of SU(3) Holonomy},
\newblock Nucl. Phys. {\bf B 268} (1986) 79-112.
\bibitem{CL}
P.  Candelas and X.  De  La Ossa, {\em  Moduli space of Calabi-Yau
Manifolds},
\newblock Nucl. Phys. {\bf B 355} (1991) 455-481.
\bibitem{CGH}
P.  Candelas, R. Grenn  and  M.  Hubsc, {\em  Connected  Calabi -Yau
Compactifications},
 (World Scientific , Singapore).
\bibitem{Sen2}  A. Sen, {\em  Heterotic string theory and Calabi Yau manifolds
 in the Green Schwarz Formalism},  Nucl. Phys. {\bf  B355} (1987) 423.
\bibitem{NSV}
K.~S. Narain, M.~H. Sarmadi, and C.~Vafa,
\newblock {\em Asymmetric Orbifolds},
\newblock Nucl. Phys. {\bf B288} (1987)551-577.
\bibitem{P} A. M. Polyakov, {\em particle spectrum in the quantum field theory},
JETP. Lett {\bf 20} (1974)494-195.
\bibitem{T}
 G.'t Hooft, Computation of the quatum effects due to a four dimensional
pseudoparticle, Phys. Rev {\bf D14} (1976)3432-3450.
\bibitem{PP}
J.~Polchinski,
\newblock {\em Dirichlet-Branes and Ramond-Ramond Charges},
\newblock Phys. Rev. Lett. {\bf 75} (1995) 4724-4727,  {\tt hep-th/9510017}.
\bibitem{W1} E. Witten, Bound States  of  Strings   and  p-Branes, Nucl.
Phys. {\bf 460} (1996) 335, {\tt  hep-th/9510135}.
\bibitem{BSV1} M. Bershadsky, V. Sadov and C. Vafa, {\em  D-srtings on D- Manifols},
Nucl. Phys. {\bf 463} (1996) 398.
\bibitem{BSV2} M. Bershadsky, V. Sadov and C. Vafa, {\em D-branes and topological
field theories},  Nucl. Phys. {\bf 463} (1996) 420.
\bibitem{DKP} M. Douglas,  D. Kobat, P. Pouliot  and  S. Shenker, {\em  D- branes
and Short Distances in  String Theory},  {\tt  hep-th/9608024}.
\bibitem{SW3}
 N. Seiberg, E.  Witten,  {\em The D1/D5 System And Singular CFT}, JHEP 9904 (1999)
{\bf 017}, {\tt hep-th/9903224}.
\bibitem{PW}
J. Pochinski and E. Witten, { \em  Evidence  for Heterotic- type I
String Duality}, Nucl. Phys. {\bf B460} (1996) 525, {\tt  hep-th/9510169}.
\bibitem{Aspin2}
P.~S. Aspinwall,
\newblock {\em Enhanced Gauge Symmetries and K3 Surfaces},
\newblock Phys. Lett. {\bf B357} (1995) 329-334, {\tt hep-th/9507012}.
\bibitem{Aspin3}
P.~S. Aspinwall, Phys. Lett {\bf 357}(1995) 329.
\bibitem{Ber}
M.~Bershadsky et~al.,
\newblock {\em Geometric Singularities and Enhanced Gauge Symmetries},
\newblock Nucl. Phys. {\bf B481} (1996) 215-252, {\tt hep-th/9605200}.
\bibitem{KMP}
S.~Katz, D.~R. Morrison, and M.~R. Plesser,
\newblock {\em Enhanced Gauge Symmetry in Type II String Theory},
\newblock Nucl. Phys. {\bf B477} (1996) 105-140, {\tt hep-th/9601108}.
\bibitem{HT2}
C.~Hull and P.~Townsend,
\newblock {\em  Enhanced  Gauges Symmetries and K3 surfaces},
\newblock  Phys. Lett.  {\bf B 347} (1995)313.

\bibitem{AFIQ}
G.~Aldazabal, A.~Font, L.~E. Ib{\'a\~n}ez, and F.~Quevedo,
\newblock {\em Chains of N=2, D=4 Heterotic/Type II Duals},
\newblock Nucl. Phys. {\bf B461} (1996) 85-100, {\tt hep-th/9510093}.\\
S.~Chaudhuri and D.~A. Lowe,
\newblock {\em Type IIA-Heterotic Duals with Maximal Supersymmetry},
\newblock Nucl. Phys. {\bf B459} (1996) 113-124, {\tt hep-th/9508144}.
\bibitem{Kachru}
S.~Kachru et~al.,
\newblock {\em Nonperturbative Results on the Point Particle Limit of N=2
  Heterotic String Compactifications},
\newblock Nucl. Phys. {\bf B459} (1996) 537-558, {\tt  hep-th/9508155}.
\bibitem{KLM}
A.~Klemm, W.~Lerche, and P.~Mayr,
\newblock {\em K3--Fibrations and Heterotic-Type II String Duality},
\newblock Phys. Lett. {\bf 357B} (1995)313-322,  {\tt hep-th/9506112}.
\bibitem{Klemm}
A.~Klemm et~al.,
\newblock {\em Self-Dual Strings and N=2 Supersymmetric Field Theory},
\newblock Nucl. Phys. {\bf B477} (1996)746-766,  {\tt hep-th/9604034}.
\bibitem{KLT}
V.~Kaplunovsky, J.~Louis, and S.~Theisen,
\newblock {\em Aspects of Duality in N=2 String Vacua},
\newblock Phys. Lett. {\bf 357B} (1995)71--75, {\tt  hep-th/9506110}.
\bibitem{OV}
H. Ooguri, C. Vafa,  {\em All Loop N=2 String Amplitudes},
  Nucl.Phys. {\bf B451} (1995)121-161, {\tt hep-th/9505183}.
\bibitem{SV}
  A. Sen, C. Vafa, {\em Dual Pairs of Type II String Compactification},  Nucl.Phys.
{\bf B455} (1995)165, {\tt hep-th/9508064}.
\bibitem{Sen4}
A. Sen, {\em  An introduction to Non-perturbative string theory},
{\tt hep-th/9802051}.
\bibitem{VW} C. Vafa and E. Witten, Nucl. Phy.  Proc.  Supp {\bf 46} (1996)225.
\end{thebibliography}
\end{document}